\documentclass[notitlepage,nobibnotes,superscriptaddress,aps,prl,twocolumn,noeprint]{revtex4-1}

\usepackage[english]{babel}
\usepackage{threeparttable}
\usepackage{graphicx}
\usepackage{caption}
\usepackage[caption=false,position=top,singlelinecheck=off,justification=raggedright]{subfig}
\usepackage {cancel}
\usepackage{caption}
\usepackage{mathtools}

\usepackage{mathdots}
\usepackage[utf8]{inputenc}
\usepackage{times}
\usepackage{newtxtext}
\usepackage{xcolor}
\usepackage{multibib}
\newcites{appx}{References}

\usepackage{mathptmx}
\usepackage{amsthm}

\usepackage{caption}
\newcommand{\TScom}[1]{}

\newcommand{\RHcom}[1]{}

\newcommand{\YNcom}[1]{}

\makeatletter
\@ifundefined{auto@bib}{}{\let\auto@bib\@empty}
\@ifundefined{write@bibliographystyle}{}{\let\write@bibliographystyle\relax}
\@ifundefined{@bibdataout@init}{}{\let\@bibdataout@init\relax}
\@ifundefined{@bibdataout@rev}{}{\let\@bibdataout@rev\relax}
\makeatother

        \usepackage[top=30mm, bottom=30mm, left=25mm, right=25mm]{geometry}
 \usepackage{amsmath}
        \usepackage{amsmath, amssymb}
	\usepackage{color}
	\usepackage{graphicx}
	\usepackage{braket}
	\usepackage[subrefformat=parens]{subcaption}
	\usepackage{txfonts}
	\usepackage{enumerate}
	\usepackage{bm}
	
 \usepackage{ragged2e}

	\definecolor{gray}{gray}{0.5}
	\definecolor{emerald}{cmyk}{1,0,0.5,0}
	\definecolor{bluegreen}{cmyk}{0.85,0,0.33,0}
	\definecolor{violet}{cmyk}{0.79,0.88,0,0}

\usepackage{soul}
\usepackage{ulem}
\usepackage{lipsum}
\usepackage{footnote}

\makeatletter
\renewcommand{\textcolor}[3][]{#3}

\renewcommand{\sout}[1]{}
\renewcommand{\st}[1]{}
\makeatother

\begin{document}
\preprint{APS/123-QED}

\title{Continuous Time Crystals as a $\mathcal{PT}$ Symmetric State and \\
the Emergence of Critical Exceptional Points}

\author{Yuma Nakanishi}
\email{nakanishi.y@phys.s.u-tokyo.ac.jp}
\affiliation{Institute for Physics of Intelligence, University of Tokyo, 7-3-1 Hongo, Bunkyo-ku, Tokyo 113-0033, Japan}

\author{Ryo Hanai}
\affiliation{Department of Physics, Tokyo Institute of Technology, 2-12-1 Ookayama, Meguro-ku, Tokyo 152-8551, Japan}

\author{Tomohiro Sasamoto}
\affiliation{Department of Physics, Tokyo Institute of Technology, 2-12-1 Ookayama, Meguro-ku, Tokyo 152-8551, Japan}

\date{\today}

\begin{abstract}

Continuous time-translation symmetry is often spontaneously broken in open quantum systems, and the condition for their emergence has been actively investigated. However, there are only a few cases in which its condition
for appearance has been fully elucidated.
In this Letter, we show that a Lindbladian parity-time ($\mathcal{PT}$) symmetry can generically produce persistent periodic oscillations in a wide class of systems.
This includes one-collective spin models, which have been studied thoroughly in the context of dissipative continuous time crystals, and spatially extended bipartite bosonic systems with conserved particle number.
By making an analogy to nonreciprocal phase transitions, we demonstrate that a transition point from the dynamical phase is associated with spontaneous PT symmetry breaking that typically corresponds to a \textit{critical exceptional point}.
Interestingly, the periodic orbits in the PT-symmetric phase are found to be center-type, implying an initial state dependence.
These results are established by proving that the Lindbladian $\mathcal{PT}$ symmetry at the microscopic level implies a nonlinear PT symmetry, and by performing a linear stability analysis near the transition point. This research will further our understanding of novel non-equilibrium phases of matter and phase transitions with spontaneous anti-unitary symmetry breaking.

\end{abstract}

\maketitle

\textit{Introduction}. ---
Exploration of phases of matter unique to systems out of equilibrium is an important problem in non-equilibrium statistical physics.
A paradigmatic example of such nonequilibrium exotic states of matter is a continuous time crystal $\text{\cite{Wilczek, Sacha, Else}}$,
which spontaneously breaks the continuous time-translation symmetry into a discrete one.
This has been proven impossible in equilibrium systems {(with short range interaction)} \cite{Watanabe}, but it is possible to exist out of equilibrium.

Those in open systems, called \textit{dissipative continuous time crystals (DCTCs)},
have been shown to arise in various quantum systems.
This includes driven-dissipative collective spin (that consists of all-to-all coupled two-level systems) \cite{Iemini, Piccitto, dos, Buonaiuto}, bosonic \cite{Minganti4, Lled2, Cabot, Li}, fermionic \cite{Booker}, and spin-1/2 systems \cite{Passarelli, Ya-Xin, Yang}, and have recently been observed experimentally \cite{Kongkhambut, Wu, Jiao, Chen, Greilich}.
Many of these, in the classical limit, correspond to the limit cycles or closed orbits in nonlinear dynamical systems, which have traditionally been studied in various classical contexts, including chemistry and biology \cite{Kuramoto}.
Curiously, for collective spin systems, it has been suggested that the emergence of these \textit{macroscopic} oscillations is related to some set of \textit{microscopic} symmetry such as ${\mathbb{Z}}_2$ \cite{dos, Piccitto} or parity-time ($\mathcal{PT}$) symmetry \cite{Nakanishi2}.
However, they are based on explicit calculations for several concrete models, and
the reasoning for such symmetry requirements and their exact role is yet to be elucidated.

One of the reasons for such difficulties is that, in the GKSL equation, extracting the physical consequences of symmetries is not always trivial.
For unitary symmetries, such as $U(1)$-symmetry, they indicate the existence of a conserved charge \cite{Buca2} or the occurrence of phase transition with their symmetry breaking in the steady-state \cite{Minganti, Kessler}.
However, for \textit{anti}-unitary symmetry, such as $\mathcal{PT}$ symmetry, it is difficult to extract such conclusions.
Technically, this is due to the property that the operator corresponding to this symmetry
does not commute with the generator of the dynamics (i.e., the non-shifted Lindbladian), in contrast to conventional symmetries.

In this Letter, despite the challenges mentioned above, we show that a $\mathcal{PT}$ symmetry at a \textit{microscopic} level generically implies the emergence of DCTCs
in a wide class of systems.
Specifically, we establish that bipartite bosonic systems with $\mathcal{PT}$ symmetry and conserved particle numbers,
which includes
spatially extended systems and those that can be mapped to collective spin systems, generically exhibit time-dependent phases that, interestingly, have initial-state dependence.
This is shown by connecting the two different notions of $\mathcal{PT}$ symmetries for microscopic and macroscopic quantities developed in different contexts.
We prove that, if the GKSL equation has a \textit{Lindbladian} $\mathcal{PT}$ (L-$\mathcal{PT}$) symmetry at the microscopic level \cite{Huber2, Nakanishi2}, its mean-field equation
has \textit{nonlinear}
PT (n-PT)
symmetry \cite{Fruchart, Konotop} and its PT-symmetric solution generically
exhibits
persistent oscillations.
We demonstrate that transition points from the dynamical phase are associated with spontaneous n-PT symmetry breaking.
The transition points are typically marked by so-called critical exceptional points (CEPs), where the criticality occurs by the coalescence of collective excitation mode to a zero mode \cite{Fruchart, Hanai, Hanai2, You, Saha, Zelle, Suchanek, Chiacchio, Nadolny}.
Finally, we demonstrate our argument for a one-collective spin model which can be mapped to two-boson systems with conserved particle number.

Our work provides an intriguing connection to a class of nonequilibrium phase transitions called
\textit{nonreciprocal phase transitions} \cite{Fruchart, You, Saha}, which are also characterized by CEPs.
In an active (classical) system where the detailed balance is broken, constituents do not necessarily satisfy the action-reaction symmetry (e.g., particle A attracts particle B but B repulses A).
In such systems, it was found that a phase transition from a static to a time-dependent phase occurs, where the latter corresponds to a phase where collective degrees of freedom exhibits a persistent many-body chase-and-runaway motion \cite{Fruchart, You, Saha}.

Interestingly, the symmetry (and its breaking) of bipartite bosonic systems with conserved particle number
considered in this work is \textit{conceptionally} similar
to those of nonreciprocal phase transitions,
but with an important difference
originating from their physical context \cite{ptantipt}.
In particular, while the former breaks the PT symmetry, the latter breaks the \textit{anti}-PT symmetry.
This subtle difference leads to significant physical consequences.
For our systems,
linear stability analysis reveals that all the physical PT-symmetric fixed points are center types (also called neutrally stable fixed points), implying the presence of
persistent periodic closed orbits.
Our analysis also shows that the nature of the transition in DCTCs
has an unusual property compared to the conventional symmetry breaking that
a pair of stable and unstable solutions appear instead of steady-state degeneracy in a symmetry-broken phase.
These are in contrast to nonreciprocal systems, where the anti-PT-symmetric (broken) phase has a unique (two) stable fixed point (limit cycles) as their stationary states \cite{Fruchart, You, Saha}.

\textit{Lindbladian $\mathcal{PT}$ symmetry}. ---
In open quantum systems where the evolution of states is completely positive and trace-preserving (CPTP) Markovian, the time evolution of the density matrix $\rho(t)$ is described by the GKSL equation $\text{\cite{Lindblad, GKS}}$,
$\partial_{t}\rho=\hat{\mathcal{L}}\rho=-i[H,\rho]+\sum_{\mu}\hat{\mathcal{D}}[L_{\mu}]\rho$, where $H$ is the Hamiltonian, $L_{\mu}$ is the Lindblad operator, and $\hat{\mathcal{L}}$ is the Lindbladian. Here, the dissipation superoperator $\hat{\mathcal{D}}[L]$ is defined as $\hat{\mathcal{D}}[L]\rho=2L\rho L^{\dagger}-L^{\dagger}L\rho-\rho L^{\dagger}L.$

We say that $\hat{\mathcal L}$ is L-$\mathcal{PT}$ symmetric when the following relation is satisfied~\cite{Nakanishi1, Nakanishi2, Huber1, Huber2}:
\begin{align}
\label{HuberPT}
\hat{\mathcal{L}}[\mathbb{PT}(H); \mathbb{PT}(L_{\mu}),\mu=1,2,\cdots]
=\hat{\mathcal{L}}[H;L_\mu,\mu=1,2,\cdots],
\end{align}
{with $\mathbb{PT}(O):=PT\,O^{\dagger}(PT)^{-1}$.
Here, $P$ is taken to be a self-inverse permutation operator, satisfying
$P^2=1$ and $P=P^\dagger$.
The time-reversal operator $T$ is anti-unitary and commutes with $P$, i.e., $[P,T]=0$.}

We remark that
our L-$\mathcal{PT}$ symmetry \eqref{HuberPT} is different from the $\mathcal{PT}$ symmetry for a shifted Lindbladian analyzed in Refs.~\cite{Prosen3, Prosen4,Sa, Huybrechts}
defined as anti-pseudo Hermicity \cite{App,Prosen3}.
The consequence of L-$\mathcal{PT}$ symmetry \eqref{HuberPT} considered here turns out to be very different: as shown below, the $\mathcal{PT}$ symmetry and its spontaneous breaking implies the emergence of phase transitions to \textit{steady state} properties (including stationary oscillation) in the thermodynamic limit, not only their spectral properties (as shown in Table A.1 in Supplemental Material (SM) Sec. A \cite{supsup}).

\textit{Nonlinear PT symmetry for mean-field equation}.---
In the following, we will show that the L-$\mathcal{PT}$ symmetry~\eqref{HuberPT}
in a dissipative bosonic system on a lattice with $l$ sites
puts a strong constraint on its mean-field dynamics.
We define
the parity operator $P$
as a permutation acting on the Fock state $\ket{n_1,n_2,...,n_l}$ as $P\ket{n_1,n_2,...,n_l}=\ket{n_{\sigma(1)},n_{\sigma(2)},...,n_{\sigma(l)}}$.
Here, $\sigma:\{1,2,...,l\}\to\{1,2,...,l\}$ is a permutation that satisfies $\sigma^2=1$ and $n_j:=a_j^\dagger a_j$ denotes the particle number at the $j$-th site with the bosonic annihilation (creation) operator $a_j$ ($a_j^\dagger$).
The parity operator acts on the bosonic annihilation operators as $Pa_jP^{-1}=a_{\sigma(j)}$. The time-reversal operator $T$ {is taken to be} complex conjugation $i\to-i$.

Below, we restrict ourselves to the case where the total particle number $N:=\sum_j n_j$ is conserved and very large $N\gg 1$ while keeping the number of sites $l(\ll N)$ fixed.
We rescale the bosonic operators as $b_i:=a_i/\sqrt{N}$. {Under appropriate scalings} of a Hamiltonian and a set of Lindblad operators (See SM Sec. B for detail \cite{supsup}), the mean-field approximation, replacing bosonic operators with the quantum expectation value $\braket{b_i}$ after normal ordering, becomes exact if an initial state is a tensor product of coherent states in the large-$N$ limit \cite{supsup, Souza, Carollo}. Here, $\braket{\cdot}$ denotes Tr$[\ \cdot\ \rho]$.

{The mean-field equation for local observables $q_i:=\lim_{N\to\infty}\braket{b_i}$
is written in the form of nonlinear dynamical systems
$i\partial_t{\bf{q}}={\bf{f}}({\bf{q}})$
with ${\bf{q}}:=(q_1,q_2,..,q_l)^T\in\mathbb{C}^{l}$ and a nonlinear function
${\bf{f}}=(f_1, f_2,...,f_l): \mathbb{C}^{l}\to \mathbb{C}^{l}$.}
To put constraints on their form,
we find it useful to define the notion of
symmetry of nonlinear dynamical systems.
The nonlinear dynamical system is said to be nonlinear X symmetric (n-X symmetric) if $\tilde{X}{\bf{f}}({\bf{q}})={\bf{f}}({\tilde{X}\bf{q}})$, where $\tilde{X}$ is
a unitary or anti-unitary operator. (Symmetry operations in nonlinear dynamical systems are denoted with tildes to distinguish them from those of operators.)

In this Letter,
we will be concerned with the case of $\tilde {X}=\tilde {P} \tilde {T}$,
where we define a parity operator as
$\tilde{P}{\bf{f}}=(f_{\sigma(1)},f_{\sigma(2)},...,f_{\sigma(l)})^T$
and the time-reversal operation
as $\tilde T {\bf{f}}(t)={\bf f}^*(-t)$, so that
the n-PT symmetry
$\tilde{P}\tilde{T}{\bf{f}}({\bf{q}}(t))={\bf{f}}(\tilde{P}\tilde{T}{\bf{q}}(t))$
reads
\begin{eqnarray}
\label{nonlinearPT}
\tilde{P}{\bf{f}}^*({\bf{q}}(t))
={\bf{f}}({\tilde{P}\bf{q}^*}(t)).
\end{eqnarray}
{From the mean-field equation,
one finds that
if ${\bf{q}}(t)$ is a solution, then $\tilde{P}\tilde{T}{\bf{q}}(t)$ is also a solution. }
With the above preparations, we are now ready to state the main theorem of this paper:\\

\noindent\textbf{Theorem}
\textit{For a dissipative bosonic system with conserved particle number and L-$\mathcal{PT}$ symmetry ($\ref{HuberPT}$), its mean-field {equation} {possesses an} n-PT symmetry ($\ref{nonlinearPT}$) {in the large total particle number limit}.
}\\

\noindent We provide a proof in SM Section B \cite{supsup}.

\textit{Linear stability analysis}.---
So far, we have established that when the GKSL equation has the L-$\mathcal{PT}$ symmetry (Eq.~\eqref{HuberPT}), the corresponding order parameter dynamics for a large-$N$
have an n-PT symmetry (Eq.~\eqref{nonlinearPT}).
For conventional symmetries, the steady state (or thermal state in equilibrium) may spontaneously break.
For example, a paramagnetic phase in the Ising model with a $\mathbb{Z}_2$ symmetry may destabilize into a ferromagnetic phase that spontaneously breaks the $\mathbb{Z}_2$ symmetry.

We show below that a similar destabilization of PT-symmetric state in Eq.~\eqref{nonlinearPT} may occur, implying a spontaneous n-PT symmetry breaking in the steady state.
We first define what we mean by spontaneous n-PT symmetry breaking.
{Here, the PT symmetry of a fixed point ${\bf{q}}$ is said to be unbroken if ${\bf{q}}=\tilde{P}{\bf{q}^*}$; otherwise, the symmetry is spontaneously broken.}

To be concrete,
we focus on a bipartite system with a uniform solution, the minimal case with spontaneous n-PT symmetry breaking among the dissipative bosonic systems.
{For bipartite systems with sublattices $A$ and $B$, the uniform solution is characterized by two order parameters, $q_A$ and $q_B$, which denote the uniform values of $q_i$ on sublattices $A$ and $B$, respectively.}
Since the bosonic particle number is conserved, that is, the system has (strong) $U(1)$ symmetry, the mean-field equation is independent of each phase $\theta_{A},\theta_{B}$ in the polar coordinates {$q_{A}=r_{A}e^{i\theta_{A}}$, $q_{B}=r_{B}e^{i\theta_{B}}$.}
Together with the property that the total number is conserved $r_{\rm A}^2+r_{\rm B}^2={\rm const.}$,
the mean-field equation
can be specified to the form
\begin{align}
\label{rarbthe}
\partial_{t}{\bf{q}^\prime}=\begin{pmatrix}
G(r_A,r_B,\Delta\theta)/r_A\\
-G(r_A,r_B,\Delta\theta)/r_B \\
H(r_A,r_B,\Delta\theta)\\
\end{pmatrix},
\end{align}
with ${\bf{q}}^\prime:=(r_A,r_B,\Delta\theta)^T\in\mathbb{R}^{3}$and the phase difference $\Delta\theta:=\theta_{A}-\theta_{B}$.
Here, the function $G$ ($H$) is symmetric (anti-symmetric) for the exchange of amplitudes $r_A$, $r_B$. As expected by Theorem, Eq.~($\ref{rarbthe}$) also has an n-PT symmetry where the parity operator satisfies the relation $\tilde{P}{\bf{q}^\prime}=(r_B,r_A,\Delta\theta)^T$.

Now, let us perform the linear stability analysis around a $\mathcal{PT}$-symmetric fixed point by calculating the Jacobian matrix $J$ with $\partial_t\delta{\bf{q}^\prime}=J\delta{\bf{q}^\prime}$, where $\delta{\bf{q}^\prime}:=(\delta r_B,\delta \Delta\theta)^{T}$ represents the fluctuation vector around the fixed point.
Here, we have eliminated the $r_A$ component
using the property that
the total particle number is conserved.
From Eq.~($\ref{rarbthe}$), the Jacobian around the PT-symmetric fixed point with $r_A=r_B$ can be simplified to the form
\begin{align}
\label{LL}
J=\begin{pmatrix}
0&\alpha\\
\beta&0\\
\end{pmatrix},
\end{align}
where $\alpha:=\left.-\partial_{\Delta\theta}G/r_B\right|_{ss},\ \beta:=(\left.\partial_{r_B}H-\partial_{r_A}H)\right|_{ss}\in\mathbb{R}$.
Here, quantities with ``${}|_{ss}$" describe the value at the fixed point.

The excitation spectra and modes of Eq.~($\ref{LL}$) are given by $\lambda=\pm\sqrt{\alpha\beta}$ and $\delta{\bf{q}^\prime}=(\pm\sqrt{\alpha/\beta},1)^T$.
Physically, the real part of the eigenvalue $\lambda$ characterizes the growth rate of the fluctuations, while the imaginary part represents the oscillation frequency.
For the PT-symmetric solution to be physical, $\alpha\beta$ must be non-positive, since otherwise ($\alpha\beta>0$) the eigenvalue would have a positive real part, implying the diverging growth of fluctuation and, therefore, the solution is unstable.

In the physical case of $\alpha\beta<0$, purely imaginary eigenvalues emerge.
This fixed point type is called \textit{center} \cite{Strogatz}.
The finite imaginary part implies the occurrence of
periodic closed orbits around the fixed point,
while the vanishing real part shows that the oscillation is persistent.
As the decay is absent, this implies that the resulting orbits are initial state dependent.
It is interesting that despite the presence of dissipation, the PT symmetry of the system ensures the presence of such marginal orbits.

Starting from this PT-symmetric state $\alpha\beta<0$, one may move microscopic parameters in the Hamiltonian or Lindblad operator
such that $|\alpha\beta|$ decreases,
until it reaches a critical point $\alpha\beta = 0$.
In this situation, the frequency $\omega=\sqrt{|\alpha\beta|}$ vanishes, signaling the divergence of a timescale \cite{Note1}.
Interestingly, this critical point is generically characterized by the \textit{coalescence} of the eigenmodes to a zero mode \cite{Fruchart, Hanai, Hanai2, You, Saha, Zelle, Suchanek, Chiacchio, Nadolny},
which is called a critical exceptional point (CEP) in the literature \cite{Hanai2, Zelle}.
A notable exception is when $\alpha$ and $\beta$ simultaneously vanish: in this case, all the elements in the Jacobian $\eqref{LL}$ are zero, leading to having a complete basis and hence no CEP. (We will see an example of this special case just before the conclusion.)

One may also perform a similar linear stability analysis to the PT-broken fixed points as well.
It can be shown generically that the fixed points come in pairs, where
one is stable and the other is unstable in the physical case (see SM Sec. C \cite{supsup}).
We remark that this is different from conventional phase transitions, where a pair of stable fixed points appears in a symmetry-broken phase \cite{Minganti}.

In summary, by combining Theorem with linear stability analysis we have shown that the L-$\mathcal{PT}$ symmetry ($\ref{HuberPT}$) can generically produce persistent periodic oscillations for bipartite bosonic models with conserved particle number. Moreover, we have revealed that a pair of stable and unstable fixed points emerges if a PT symmetry of solutions is broken and the transition point is typically a CEP.

At a glance, this is similar to the PT transitions of non-Hermitian Hamiltonians, which are also associated with exceptional points $\text{\cite{MostafazadehA1, Bender, Bender2}}$.
However, there is a fundamental difference: the non-Hermitian PT symmetry breaking is a \textit{spectral} transition while our n-PT symmetry breaking is a transition of a \textit{steady state} (which is closer to the conventional notion of phase transition).
As a result, the exceptional points that mark our transition are \textit{critical} (i.e. the damping rate vanishes), while the former is generically not \cite{MostafazadehA1, Bender, Bender2} (as shown in Table A.1 in SM Sec. A \cite{supsup}) \cite{Note2}.

\textit{One-collective spin models.---}
A representative example to which
our framework can be applied
is the one-collective spin
~\cite{Hannukainen, Ribeiro, Carmichael, Ferreira}.
(For a concrete example of a spatially extended bosonic system, see SM Sec. D.5 \cite{supsup}.)
This class of systems has been discovered to exhibit DCTCs \cite{Iemini, Piccitto, dos, Buonaiuto}
and has been extensively studied since then.
Since it is known that one-collective spin can be mapped to a two-boson particle-number conserving system via the Schwinger boson transformation \cite{Pires},
our conclusions for bipartite bosonic systems with an L-$\mathcal{PT}$ symmetry above can be directly applied when the collective spin has the L-$\mathcal{PT}$ symmetry.

\begin{figure}[t]
   \vspace*{-0.7cm}
     \hspace*{-4.7cm}
\includegraphics[bb=0mm 0mm 90mm 150mm,width=0.46\linewidth]{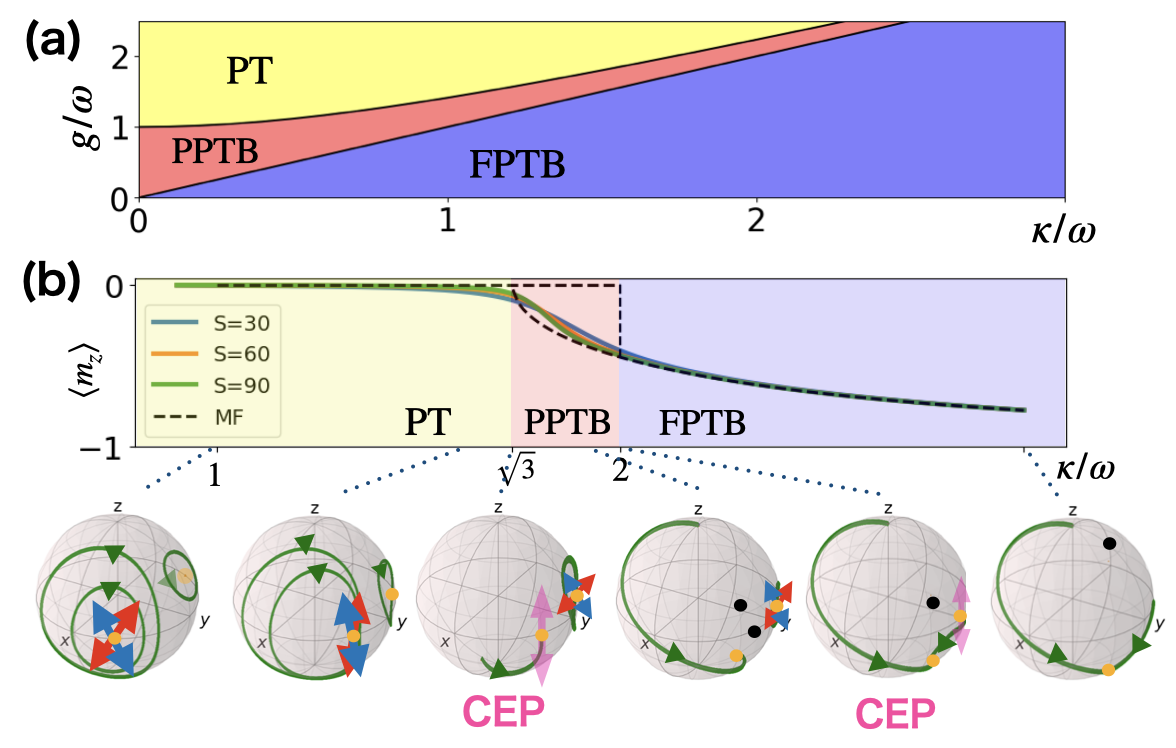}
\caption{\small\justifying (a) The phase diagram of the generalized DDM. The yellow, blue and red regions can be classified based on the PT symmetry of the fixed points and are referred to as PT-symmetric, fully PT-broken (FPTB), and partially PT-broken (PPTB) phases, respectively. In these regions, there are two PT-symmetric, two PT-broken, and both PT-symmetric and PT-broken fixed points, respectively. (b) Top: The normalized magnetization $\braket{m_{z}}$ in the stationary state $\lim_{t\to\infty}\rho(t)$ for finite $S$ and mean field solution ($S=\infty$) with $\omega=1,g=2$. Bottom: Mean-field trajectory of dynamics (green arrow), stable (orange) and unstable (black) fixed points, and the representative collective excitation modes (blue and red or pink arrows). Here, the components of collective excitation modes are generally complex numbers, so we take absolute values. The transition points are CEPs.}
    \label{lpt}
\end{figure}

Let us first consider
the generalized Driven Dicke model (DDM) described by
the Hamiltonian $H=S(2gm_{x}+\omega m_z^2)$,
where the transverse magnetic field with strength $g$ and
two-body interaction along the z-axis with strength $\omega$,
and the Lindblad operator representing the collective decay with rate $\kappa$, $L=\sqrt{\kappa S}m_{-}$, where $m_{\alpha}:=\sum_{i=1}^{N}\sigma_\alpha^i/S$ ($\alpha=x,y,z$) is the normalized magnetization with Pauli matrices $\sigma_\alpha$ and $m_{-}:=m_x-im_y$ denotes the lowering operator.
Here, the total spin $S:=N/2$ is conserved, where $N$ represents the number of two-level systems.
Note that
the implementation of all-to-all (or long-range) interactions has already been demonstrated \cite{Baumann, Zhang, Baumann2} and recently collective decay has also been realized in hybrid quantum systems \cite{Angerer} and cold atom systems in free space \cite{Ferioli}, indicating that these experimental setups serve as promising platforms for implementing dissipative collective spin models \cite{Note3}.

This model has a L-$\mathcal{PT}$ symmetry $\eqref{HuberPT}$ \cite{Nakanishi2}, where the parity operator is a $\pi$-rotation of collective spin around the $x$-basis {$P=\prod_i\sigma^i_x$}.
As stated above, it
can be mapped to a two bosonic system by the Schwinger boson transformation, $m_-=a b^\dagger/(2S),\ \ m_z=(a^\dagger a-b^\dagger b)/(2S)$, $S=(a^\dagger a+ b^\dagger b)/2$, where $a,b$ are bosonic annihilation operators.

The time evolution in the large $S$ limit is given by \cite{Souza, Carollo}
{\begin{align}
\label{timematrix1}
i\partial_{t}{\bf{M}}
=
2i\begin{pmatrix}
-\omega YZ+\kappa XZ\\
\omega XZ-gZ+\kappa YZ\\
gY-\kappa(1-Z^2)\\
\end{pmatrix},
\end{align}
with ${\bf{M}}=(X,Y,Z)^T:=\lim_{S\to\infty}(\braket{m_{x}},\braket{m_{y}},\braket{m_{z}})^T$}.
As expected, the dynamical system ($\ref{timematrix1}$) also exhibits the n-PT symmetry ($\ref{nonlinearPT}$) with $\tilde{P}=diag(1,1,-1)$.

Figure~\ref{lpt}(a) shows the phase diagram of this model,
which consists of three regimes.
The yellow region
is a PT-symmetric regime with
two stable fixed points ($Z=0$, i.e. $r_A=r_B$) that respect PT symmetry,
while the blue region exhibits two PT-broken fixed points ($Z \ne 0$).
In the red region, we find that a pair of stable and unstable fixed points of both PT-symmetric and PT-broken fixed points coexist.
The PT-symmetric phase has eigenvalues $\lambda_\pm$ of the Jacobian that are pure imaginary,
$\lambda_\pm=\pm i\sqrt{|\alpha_\pm \beta|}$
with
{$\alpha_{\pm}=-2\left. X_\pm\right|_{ss}(g\left.X_\pm\right|_{ss}-\omega)$, $\left.X_\pm\right|_{ss}=\pm\sqrt{1-(\kappa/g)^2}$},
and $\beta=2g$, implying that the orbits are marginal.
At both the phase boundaries (yellow to red and red to blue), we find that a PT-symmetric phase destabilizes at the critical exceptional point \cite{Sup2}.
All these behaviors are consistent with our general theory developed above.

Reference~\cite{Piccitto} argued, based mainly on numerical results,
that ${\mathbb{Z}}_2$ symmetry of the Hamiltonian part may be a necessary condition.
Our symmetry-based theory clearly shows that the condition for the emergence of DCTC and its transitions should be understood instead in terms of $\mathcal{PT}$ symmetry.
Indeed, in the model studied in Ref.~\cite{Buonaiuto}, the Lindbladian (Hamiltonian does not) have an L-$\mathcal{PT}$ ($\mathbb{Z}_2$) symmetry but exhibits a DCTC, consistent with our theory [See SM Sec. D.4 \cite{supsup}].

As demonstrated above, there are (infinitely in theory)  many one-collective spin models to which our theory can be applied, where a CEP appears at the transition point.
Next, as an example that shows a phase transition to DCTCs \textit{without} CEP, we examine the Lipkin-Meshkov-Glick (LMG) model, studied in nuclear and solid state physics \cite{Lipkin}.
This model is described by the Hamiltonian $H=gS(m_{+}^{2}+m_{-}^{2})/2$ with the collective decay $L=\sqrt{\kappa S}m_{-}$.
It also has the L-$\mathcal{PT}$ symmetry and therefore, our theory applies here as well.
The PT-symmetric and broken phase is separated by the discontinuous transition \cite{Lee}, unlike in the previous example.
In the PT-symmetric phase, the eigenvalues of the Jacobian $\lambda=\sqrt{\alpha \beta}$ are pure imaginary as expected, where
{$\alpha=-2gX|_{ss} +2\kappa Y|_{ss} ,\ \ \beta=\frac{4}{X|_{ss}}(X^2|_{ss}-Y^2|_{ss})$} can be shown to be negative.
As detailed in SM D.3 \cite{supsup}, $\alpha$ and $\beta$ vanish \textit{simultaneously} at the transition point.
As we discussed in the paragraphs below Eq.~\eqref{LL}, in such a case, it is not associated with a CEP.
(Note that, in the generalized DDM case, $\beta=2g$ is always non-zero.)

\textit{Conclusion and discussion.}---
We have shown that the L-$\mathcal{PT}$ symmetry can generically induce spontaneous continuous time-translation symmetry breaking
in spatially extended bipartite bosonic systems with conserved number, including one-collective spin systems.
Moreover, we have revealed that transition points are associated with spontaneous n-PT symmetry breaking and typically correspond to CEPs by making an analogy to nonreciprocal phase transitions.

Since our theory is based on the symmetry principle, it can predict the emergence of DCTCs without actually
solving the problem,
irrespective of the details of the system.
Especially in collective spin systems, since dissipators that respect L-$\mathcal{PT}$ symmetry include representative dissipation processes such as pump $L=m_+$, decay $L=m_-$ and dephasing $L=m_z$, our theory suggests that the DCTC is robust against a wide range of unwanted dissipation processes.
Our theory would thus serve as a useful guide for the experimental implementation of DCTCs.

In this Letter, we focused on the PT symmetry of time-independent solutions (i.e. fixed points).
It is unclear what the role of $\mathcal{PT}$ symmetry is for time-periodic solutions.
Interestingly, we numerically find that oscillating solutions are PT-symmetric in the parameter region where a PT-symmetric fixed point is found
(see SM Sec. D \cite{supsup}),
but the general features of such solutions remain elusive, which is left to future work.

\begin{acknowledgments}
\textit{Acknowledgments}---We thank Kazuya Fujimoto, Kazuho Suzuki, Yoshihiro Michishita, Masaya Nakagawa, and Hosho Katsura for fruitful discussions.
The calculations of the magnetization and its dynamics were carried out with QuTiP \cite{Qutip}.
Y. N. also acknowledges financial support from JST SPRING, Grant No. JPMJSP2106; Tokyo Tech Academy for Convergence of Materials and Informatics; and JSPS KAKENHI Grant-in-Aid for Research Activity Start-up, Grant No. JP24K22850.
R. H. was supported by JSPS KAKENHI (Grants No. 23K19034, No. 25H01364, No. 25K00935, and No. 26H00384).
The work done by T. S. was supported by JSPS KAKENHI, Grants No. JP21H04432 and No. JP22H01143.
Part of this work was performed during the stay of T. S. and R. H. at the Isaac Newton Institute of Mathematical Sciences.
\end{acknowledgments}

\textit{Data availability}---The data that support the findings of this article are openly available in Ref.~\cite{NakanishiData2026}.

\onecolumngrid
\newpage
\large{\centerline{Supplemental Materials: \textit{Continuous time crystals as a $\mathcal{PT}$ symmetric state}}}
\large{\centerline{\textit{ and the emergence of critical exceptional points}}}
\vskip\baselineskip
\large{\centerline{Yuma Nakanishi$^{1}$, Ryo Hanai$^{2}$ and Tomohiro Sasamoto$^{2}$}}
\vskip\baselineskip
\normalsize{\centerline{\textit{$^{1}$Institute for Physics of Intelligence, University of Tokyo, 7-3-1 Hongo, Bunkyo-ku, Tokyo 113-0033, Japan}}}

\vskip\baselineskip
\normalsize{\centerline{\textit{$^{2}$Department of Physics, Tokyo Institute of Technology, 2-12-1 Ookayama, Meguro-ku, Tokyo 152-8551, Japan}}}

\makeatletter
\setcounter{NAT@ctr}{0}
\makeatother
\renewcommand{\theequation}{A.\arabic{equation} }
\setcounter{equation}{0}

\normalsize

\section{A. Comparison between non-Hermitian PT transitions and Lindbladian $\mathcal{PT}$ phase transitions}
\renewcommand{\thefigure}{A.\arabic{figure} }
\setcounter{figure}{0}
\renewcommand{\thetable}{A.\arabic{table} }
\setcounter{table}{0}

In this section, we discuss the similarities and differences between PT transitions in non-Hermitian systems and L-$\mathcal{PT}$ phase transitions (Table A.1).
A Hamiltonian $H$, in the Shr$\rm{\ddot{o}}$dinger-type equation $i\partial_t \phi=H\phi$, is said to be PT-symmetric if $H$ commutes with the combined parity and time-reversal operator,
\begin{align}
\label{PT}
[H, PT]=0.
\end{align}
If $\phi_i \propto PT\phi_i$ for all the eigenvectors $\phi_i$, the PT symmetry is said to be unbroken. Otherwise, the PT symmetry is said to be broken.

\begin{figure}[hbtp]
   \vspace*{-3.6cm}
    \hspace*{-3.1cm}
\includegraphics[bb=0mm 0mm 90mm 150mm,width=0.24\linewidth]{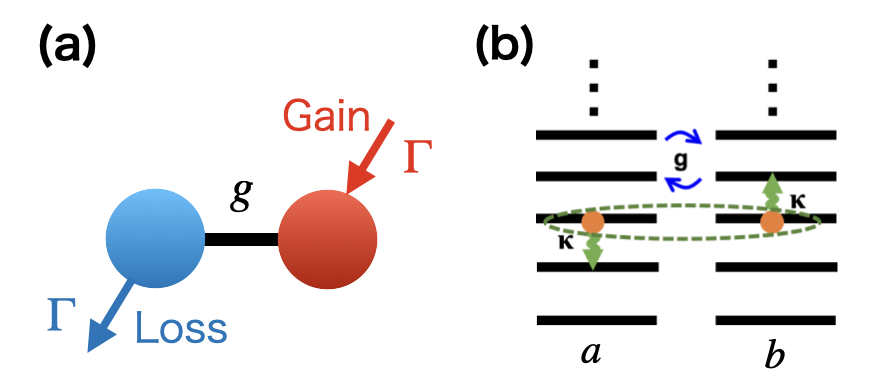}
    \caption{(a) Illustration of a typical PT-symmetric system with a balanced gain and loss. Here, $\Gamma$ and g denote the strength of gain (loss) and interaction. (b) Illustration of the DDM in the Schiwnger boson representation.}
    \label{PTpon}
\end{figure}

To see more precise property, we consider a pragmatic PT-symmetric model with balanced gain and loss, described by the non-Hermitian Hamiltonian,
\begin{align}
H=\begin{pmatrix}
-i\Gamma& g\\
g& i\Gamma\\
\end{pmatrix},
\end{align}
where $\Gamma>0$ represents the gain and loss rate, and $g\in \mathbb{R}$ is the coherent coupling strength between the two modes (Fig.$\rm\ref{PTpon}$ (a)).
The parity and time-reversal operators are defined as
$P=\begin{pmatrix}
0& 1\\
1& 0\\
\end{pmatrix}$ and $T=K$, where $K$ denotes complex conjugation.

In the weakly dissipative regime ($\Gamma<g$), gain and loss are effectively canceled by sufficiently strong interaction, \textcolor{black}{allowing the system to behave as if dissipation were absent. In fact, all the eigenvalues are real and oscillating solutions emerge.
In contrast, in the strong dissipative regime ($\Gamma>g$), the gain and loss dominate the interaction, and then a pair of complex conjugate eigenvalues appear. This signals the onset of divergent and decaying behavior over time, corresponding to stable and unstable modes.
The transition point is marked by the spontaneous breaking of the PT symmetry and coincides with an exceptional point where two eigenvectors coalesce.}

\begin{table*}[hbpt]
\centering
\small
  \begin{tabular}{|c||c|c|}  \hline
    & Non-Hermitian Hamiltonian & Lindbladian \\ \hline \hline
    Symmetry & $[H,PT]=0$ & Lindbladian $\mathcal{PT}$ symmetry \\\hline
    \begin{tabular}{c} Eigenvalues  \\ (fixed points)  \end{tabular}&  \begin{tabular}{c} Real$\to$ Complex conjugate pairs \\ (i.e., a pair of stable and unstable eigenvalues)\end{tabular} & \begin{tabular}{c} Center (i.e., purely imaginary excitation spectrum) \\ $\to$  Pairs of stable and unstable fixed points  \end{tabular}   \\ \hline
    Dynamics & Oscillation $\to$ Exponentially decay/divergence & Oscillation $\to$ Exponentially decay \\ \hline
    Symmetry breaking & PT symmetry breaking of eigenvectors & $\mathcal{PT}$ symmetry breaking of solutions in the steady state \\ \hline
   System size & Finite & Thermodynamic limit  \\ \hline
   Transition point & Exceptional point & Critical exceptional point (CEP)\\\hline
  \end{tabular}
   \caption{\small{Comparison between non-Hermitian PT transitions and  Lindbladian $\mathcal{PT}$ phase transitions.}}
\end{table*}

\textcolor{black}{We next focus on models with L-$\mathcal{PT}$ symmetry ($\ref{HuberPT}$). As discussed in the main text, if the Lindbladian has the L-$\mathcal{PT}$ symmetry ($\ref{HuberPT}$), its nonlinear mean-field equation has the n-PT symmetry in a wide class. Moreover, a dissipative phase transition occurs with spontaneous n-PT symmetry breaking, and the transition point coincides with a CEP.} In the PT-symmetric phase, persistent periodic oscillations appear, while in the PT-broken phase, a pair of stable and unstable fixed points emerges.

\textcolor{black}{Despite these parallels, a key distinction remains: non-Hermitian PT transitions are essentially spectral transitions that can occur even in a finite system, whereas their Lindbladian counterparts represent bona fide phase transitions of the steady state (including stationary oscillations) that emerge only \textit{in the thermodynamic limit}.  This contrast underscores the intrinsically many-body nature of L-$\mathcal{PT}$ phase transitions.}

A paradigmatic class that exhibits L-$\mathcal{PT}$ phase transitions is the collective spin model. For a one-collective spin, the physics can be visualized, after a Schwinger boson transformation \cite{Pires}, as a balanced gain–loss system. Introducing two bosonic modes $a,b$,
\begin{align}
\label{Schwinger}
m_x=\frac{a^\dagger b+a b^\dagger}{2S},\ \ m_y=\frac{a^\dagger b-a b^\dagger}{2iS},\ \ m_z=\frac{a^\dagger a-b^\dagger b}{2S},\ \ S=\frac{a^\dagger a+ b^\dagger b}{2},
\end{align}
the collective spin operators map onto the product of the two bosonic modes $a$ and $b$.
\textcolor{black}{The parity operator $P=\prod_i\sigma^i_x$  exchanges the two bosonic subspaces, $PaP^{-1}=b$, turning an internal spin flip into the external swap
$a\leftrightarrow b$.}
Within this picture, collective decay is represented by the composite jump operator $S_-=ab^\dagger$, which combines a gain (creation) process in mode
$b$ with a loss (annihilation) process in mode
$a$.
As an example, we describe the illustration of the DDM ($H=2gSm_x,\ L=\sqrt{\kappa S}m_-$) in the Schwinger boson representation (Fig.$\rm\ref{PTpon}$ (b)).

Finally, we remark that DCTCs originating from L-$\mathcal{PT}$ symmetry cannot appear in the effective non-Hermitian Hamiltonian obtained from the GKSL equation neglecting quantum jump terms (See $L_{\mu}^{\dagger}\rho L_{\mu}$ in Eq.~(1)). This is because this effective Hamiltonian is written in the form $H-i\sum_{\mu}L_{\mu}^{\dagger}L_{\mu}$, which can only represent loss effects. Therefore, it cannot have (active) PT symmetry, and the system decays to a steady state with no oscillations in the long time limit.

\renewcommand{\theequation}{B.\arabic{equation}}
\setcounter{equation}{0}
\textcolor{black}{\section{B. Theorem and proof of the relationship between the L-$\mathcal{PT}$ symmetry and the n-$\mathcal{PT}$ symmetry}}
\renewcommand{\thefigure}{B.\arabic{figure} }
\setcounter{figure}{0}
\begingroup
\color{black}
\textcolor{black}{Consider a dissipative bosonic system on a lattice with $l$ sites that preserves the total particle number.
A Fock state is denoted as $\ket{n_1,\dots,n_l}$, where $n_i:=a_i^{\dagger}a_i$ is the occupation number on site $i$,
and $a_i$ ($a_i^\dagger$) is the bosonic annihilation (creation) operator satisfying the canonical commutation relations $[a_i,a_j^{\dagger}]=\delta_{ij}$. The total particle number is given by $N:=\sum_i n_i$.
The time derivative of the quantum expectation value $\braket{a_i}={\rm Tr}[a_i\rho]$ is given by
\begin{align}
\label{timederi}
\partial_t \braket{a_i}={\rm Tr}[a_i \partial_t\rho],
\end{align}
where the time derivative of the density matrix is governed by the GKSL equation,
{\begin{align}
\label{gkslge2}
\partial_{t}\rho=-i[H,\rho]+\sum_{\mu}(
2L_{\mu}\rho L_{\mu}^\dagger-\{L_{\mu}^\dagger L_{\mu},\rho \})
:=\hat{{\mathcal L}}\rho
\end{align}
with the Hamiltonian $H$ and the Lindblad operator $L_{\mu}$.}
We define the mean-field approximation as {an operation that replaces}
bosonic operators $a_i$ with the quantum expectation value $\braket{a_i}$ after normal ordering, {in which all creation operators are placed to the left of all annihilation operators} (e.g. $\braket{a_i^2a_i^\dagger a_j}\simeq \braket{a_i}^*\braket{a_i}^2\braket{a_j}+2\braket{a_i}\braket{a_j}$). This is equivalent to approximating the density matrix
as a tensor product of coherent states $\rho=\prod_j \ket{\alpha_j}\bra{\alpha_j}$, where $\ket{\alpha_j}$ is a coherent state satisfying $a_j\ket{\alpha_j}=\alpha_j\ket{\alpha_j}$ with $\alpha_j\in{\mathbb{C}}$.}

\textcolor{black}{Below, we consider the large total particle number \(N\) while keeping
the number of sites $l (\ll N )$ fixed
(for collective spin systems, this
{corresponds to}
the large total spin $S$).
{We find it useful to}
introduce {normalized} bosonic operators $b_i:=a_i/\sqrt{N}$, which satisfy $[b_i,b_j^\dagger]=\delta_{i,j}/N\to0\ (N\to\infty)$.
{Since $b_i$'s commute in the limit $N\rightarrow \infty$,
this implies that}
the mean-field approximation reduces to simply replacing each $b_i$ by its quantum expectation value $\braket{b_i}$.
To keep the right-hand side of Eq.~($\rm\ref{timederi}$) remain finite in the limit $N\to\infty$,
we assume that the Hamiltonian and Lindblad operators {are extensive quantities, i.e., they} scale with the total particle number; \begin{align}
\label{scaling}
H=:N h\ \ \ \  {\rm{and}} \ \ \ \ L_\mu=:\sqrt{N}l_\mu,
\end{align}
where $h$, $l_\mu$ are intensive operators of order $\mathcal{O}(1)$, and we assume below that
they are expressed as a polynomial of finite order in normalized bosonic operators $b_i$, $b_i^\dagger$.}\\

\textcolor{black}{\noindent\textbf{Proposition.}
With the above settings, if the initial state is a tensor product of coherent states,
the time evolution of the local observable in the large-$N$ limit $q_i:=\lim_{N\to\infty}\braket{b_i}$ can be written as
\begin{align}
\label{nonlini}
i\partial_t{\bf{q}}={\bf{f}}({\bf{q}}),
\end{align}
with ${\bf{q}}:=(q_{1}, q_{2},..., q_{l})\in\mathbb{C}^{l}$ and a function ${\bf{f}}:=(f_{1},f_{2},...,f_{l})^{T}\in\mathbb{C}^{l}$ with
\begin{align}
\label{fiqq}
f_i({\bf q})
=\sum_{\substack{k_{j},m_{j}=0\\ j=1,2,...,l}}{c^{(i)}_{k_{1},m_{1},..,k_{l},m_{l}}}(q_1^{*})^{k_{1}}(q_1)^{m_{1}}\cdots(q_l^{*})^{k_{l}}(q_l)^{m_{l}},
\end{align}
where the coefficients $c^{(i)}_{k_{1},m_{1},..,k_{l},m_{l}}$ are determined by the Hamiltonian and the set of Lindblad operators.}\\

We provide the proof of Proposition in the final part of this section.\\

We now consider the Lindbladian that satisfies the L-$\mathcal{PT}$ symmetry ($\ref{HuberPT}$) {and examine their role to the meanfield equation Eq.~\eqref{nonlini}}.
Here, L-$\mathcal{PT}$ symmetry is defined in Eq. ($\ref{HuberPT}$) in the main text,
reproduced below for convenience:
\begin{align}
\label{SIeq: HuberPT}
\hat{\mathcal{L}}[\mathbb{PT}(H); \mathbb{PT}(L_{\mu}),\mu=1,2,\cdots]
=\hat{\mathcal{L}}[H;L_\mu,\mu=1,2,\cdots],
\end{align}
with $\mathbb{PT}(O):=PTO^{\dagger}(PT)^{-1}$.
Here, the parity operator $P$ is taken to be a permutation of sites
and acting on Fock states as $P\ket{n_{1},...,n_{l}}=\ket{n_{\sigma(1)},...,n_{\sigma(l)}}$, with $\sigma$ being a permutation of $l$ elements $\{1,2,...,l\}$ with $\sigma^2=1$.
The time-reversal operator $T$ is taken to be complex conjugation $T=K$, and therefore, these operators satisfy $P^{\dagger}=P^{-1}=P$, $T^{\dagger}=T^{-1}=T$, $P^2=T^2=1$, and $[P,T]=0$.
Their actions on the bosonic annihilation operators read $Pa_{i}P^{-1}=a_{\sigma(i)}$ and $Ta_{i}T^{-1}=a_{i}$.

With these preparations, the theorem stated in the main text is now presented in a slightly more precise form.\\

\noindent \textbf{Theorem.} \textit{For a dissipative bosonic system with conserved particle number and L-$\mathcal{PT}$ symmetry (\ref{SIeq: HuberPT}),
the equation of motion of the meanfield
in the large total particle number limit
$i\partial_t {\bf q}={\bf f}({\bf q})$
possesses an n-PT symmetry,
i.e.,
\begin{align}
    \label{SIeq: nonlinearPT}
    {\bf f}(\tilde P \tilde T {\bf q})
    = \tilde P \tilde T
    ({\bf f}({\bf q})),
\end{align}
if the initial condition is a tensor product of coherent states.
Here, we define a parity operation as
$\tilde{P}{\bf{f}}=(f_{\sigma(1)},f_{\sigma(2)},...,f_{\sigma(l)})^T$
and the time-reversal operation
as $\tilde T {\bf{f}}(t)={\bf f}^*(-t)$.
}
\\

\begin{proof}

The central quantity of interest is the {right-hand side} of the Heisenberg equation of motion
\begin{align}
    i\partial_t \langle b_i\rangle
    =\langle i \hat{{\mathcal L}}^\dagger b_i\rangle,
\end{align}
or more specifically, the quantity $i \hat{{\mathcal L}}^\dagger b_i$.
We denote this quantity as
\begin{align}
    \label{SIeq: iL^dagger b_i}
    i \hat{{\mathcal L}}^\dagger b_i
    :=f_i (b_1, ..., \textcolor{black}{b_l}).
\end{align}

It will be shown below that the L-$\mathcal{PT}$ symmetry implies that $i\hat{{\mathcal L}}^\dagger b_i$
and its PT-transformed quantity $\hat{\mathcal P}\hat{\mathcal T}i\hat{{\mathcal L}}^\dagger (\hat{\mathcal P}\hat{\mathcal T})^{-1} b_i$
asymptotically become equal in the large $N$ limit, i.e.,
\begin{align}
    \label{SIeq: PT symmetry in large N}
    [i\hat{{\mathcal L}}^\dagger, \hat{\mathcal P}\hat{\mathcal T}]b_i
    =O(N^{-1}),
\end{align}
where the PT-symmetry transformation is defined as $\hat{\mathcal P}\hat{\mathcal T}(O)=PT O (PT)^{-1}$.
(Note the subtle difference to the PT-transformation ${\mathbb P}{\mathbb T}$ defined in the main text, ${\mathbb P}{\mathbb T}(O)=PT O^\dagger (PT)^{-1}$.)
Since the PT-transformed quantity reads,
\begin{align}
    \label{SIeq: PT transformed iL^dagger b_i}
    \hat{\mathcal P}\hat{\mathcal T}i\hat{{\mathcal L}}^\dagger (\hat{\mathcal P}\hat{\mathcal T})^{-1} b_i
    &= \hat{\mathcal P}\hat{\mathcal T}i\hat{{\mathcal L}}^\dagger
    b_{\sigma(i)}
    \nonumber\\
    &=\hat{\mathcal P}\hat{\mathcal T}
    f_{\sigma(i)}(b_1, ..., \textcolor{black}{b_l})
    \nonumber\\
    &=f^*_{\sigma(i)}(b_{\sigma(1)}, ..., \textcolor{black}{b_{\sigma(l)}}),
\end{align}
Eq.~\eqref{SIeq: PT symmetry in large N} gives the constraint
\begin{align}
    f_i(b_1, \cdots, b_N)
    =f^*_{\sigma(i)}(b_{\sigma(1)}, \cdots, \textcolor{black}{b_{\sigma(l)}})
\end{align}
for L-$\mathcal{PT}$-symmetric systems in the large $N$ limit.
Here, in the first equality of Eq.~\eqref{SIeq: PT transformed iL^dagger b_i}, we used the property $\hat{{\mathcal P}}\hat{{\mathcal T}}=(\hat{{\mathcal P}}\hat{{\mathcal T}})^{-1}$.
Using Proposition above that the creation/annihilation operators in $f_i$ can be replaced with its mean value $q_i=\langle b_i\rangle$ in the large $N$ limit, we obtain
\begin{align}
    \braket{f_i(b_1, \cdots, b_N)}
    =f_i(q_1, \cdots, \textcolor{black}{q_l})
\end{align}
and hence
\begin{align}
    f_i(q_1, \cdots, \textcolor{black}{q_l})
    =f^*_{\sigma(i)}(q_{\sigma(1)}, \cdots, \textcolor{black}{q_{\sigma(l)}})
    =\tilde P\tilde T(f_i(\tilde P\tilde T ({\bf q})).
\end{align}
Applying $\tilde P\tilde T$ to both sides gives the desired result (Eq.~\eqref{SIeq: nonlinearPT}.)

We prove below Eq.~\eqref{SIeq: PT symmetry in large N}.
The explicit form of Eq.~\eqref{SIeq: iL^dagger b_i} reads,
\begin{align}
i\hat{\mathcal{L}}^\dagger b_i=-N[h,b_i]+iN\sum_\mu[l_\mu^\dagger,b_i]l_\mu-l_\mu^\dagger[l_\mu,b_i].
\end{align}
Using $\hat{\mathcal{P}}\hat{\mathcal{T}}b_i=b_{\sigma(i)}$,
\begin{align}
\label{ilpt}
i\hat{\mathcal{L}}^\dagger (\hat{\mathcal{P}}\hat{\mathcal{T}}b_i)=-N[h,b_{\sigma(i)}]
+iN\sum_\mu
\Big([l_\mu^\dagger,b_{\sigma(i)}]l_\mu-l_\mu^\dagger[l_\mu,b_{\sigma(i)}]
\Big)
\end{align}
On the other hand, we find
\begin{align}
\label{ptil}
\hat{\mathcal{P}}\hat{\mathcal{T}}(i\hat{\mathcal{L}}^\dagger b_i)=-N[
\mathbb {PT}(h)
,b_{\sigma(i)}]
-iN\sum_\mu
\Big(
[\mathbb{PT}({l_\mu}),b_{\sigma(i)}]\mathbb{PT}(l_\mu^\dagger)-\mathbb{PT}(l_\mu)[\mathbb{PT}(l_\mu^\dagger),b_{\sigma(i)}]
\Big)
\end{align}

The L-$\mathcal{PT}$ symmetry \eqref{SIeq: HuberPT} implies that replacing the Hamiltonian and jump operators to their PT-transformed ones
\begin{align}
\label{SIeq: LPT transformation}
h\rightarrow \mathbb {PT}(h), \qquad
\{ l_\mu\}\rightarrow \{\mathbb{PT}(l_\mu)\},\quad \mu = 1,2,,...
\end{align}
leaves the system invariant \cite{Note4}.
Therefore, for L-$\mathcal{PT}$-symmetric systems, Eq.~\eqref{ptil} can be simplified to
\begin{align}
\label{ptil2}
\hat{\mathcal{P}}\hat{\mathcal{T}}(i\hat{\mathcal{L}}^\dagger b_i)
&=-N[h,b_{\sigma(i)}]-iN\sum_\mu
\Big([{l_\mu},b_{\sigma(i)}]l_\mu^\dagger-l_\mu[l_\mu^\dagger,b_{\sigma(i)}]
\Big)
\nonumber\\
&=-N[h,b_{\sigma(i)}]
+iN\sum_\mu
\Big(
l_\mu[l_\mu^\dagger,b_{\sigma(i)}]
-[{l_\mu},b_{\sigma(i)}]l_\mu^\dagger
\Big).
\end{align}
Subtracting Eq.~\eqref{ptil2} from Eq.~\eqref{ilpt} gives,
\begin{align}
    [i\hat{{\mathcal L}}^\dagger, \hat{\mathcal P}\hat{\mathcal T}]b_i
    &=iN\sum_\mu
    \Big([[l_\mu^\dagger,b_{\sigma(i)}],l_\mu]-[l_\mu^\dagger,[l_\mu,b_{\sigma(i)}]]
    \Big),
    \nonumber\\
    &=O(N^{-1})
\end{align}
\textcolor{black}{where in the second line we used the property that double commutators introduce factors proportional to $N^{-2}$, under the assumption that the intensive operators $l_\mu$ are finite-degree polynomials in the rescaled bosonic operators. This yields the desired Eq.~\eqref{SIeq: PT symmetry in large N}.}

This completes the proof of the theorem.
\end{proof}

\begin{proof}[Proof of Proposition]
With the scalings of the Hamiltonian and the set of Lindblad operators given in Eq.~\eqref{scaling},
the GKSL master equation takes the form
\begin{align}
\label{GKSL8}
\partial_{t}\rho=-iN[{h},\rho]+N\sum_{\mu}\hat{\mathcal{D}}[{l_{\mu}}]\rho.
\end{align}
Using the GKSL equation~($\rm\ref{GKSL8}$) together with Eq.~($\rm\ref{timederi}$), the time evolution of the rescaled bosonic operators can be expressed as
\begin{align}
\label{b21}
i\frac{d}{dt}
\braket{b_i}
&=\braket{i\hat{{\mathcal L}}^\dagger b_i}
\nonumber\\
&=N\left\langle [b_{i},h]+i\sum_{\mu}\left([l_{\mu}^{\dagger},b_{i}]l_{\mu}-l_{\mu}^{\dagger}[l_{\mu},b_{i}]\right)\right\rangle.
\end{align}
The right-hand side of ($\rm\ref{b21}$) can be rewritten as a sum of products of bosonic operators:
\begin{align}
\label{a61}
i\frac{d}{dt}
{\braket{b_i}}
=\sum_{\substack{k_{j},m_{j}=0\\ j=1,2,...,l}}{c^{(i,N)}_{k_{1},m_{1},..,k_{l},m_{l}}}\braket{(b_{1}^{\dagger})^{k_{1}}(b_{1})^{m_{1}}\cdots (b_{l}^{\dagger})^{k_{l}}(b_{l})^{m_{l}}},
\end{align}
where a coeffcient $c^{(i,N)}_{k_{1},m_{1},..,k_{l},m_{l}}\in\mathbb{C}$ depends on $N$, but it is at most of order $\mathcal{O}(1)$.
Here, we used the fact that the factor $N$ on the right-hand side of Eq.~($\rm\ref{b21}$) cancels with $1/N$, which arises from the commutation relation $[b_i,b_j^\dagger]=\delta_{i,j}/N$.
(Note that all terms in the right hand side of Eq.~\eqref{b21} are proportional to the commutation relation involving $b_i$.)

We now take the large-$N$ limit of Eq.~($\rm\ref{a61}$).
With the assumption that the initial state is a tensor product of coherent states,
we show below that the average over the product of the creation/annihilation operators can be replaced with the product of their average, i.e.,
\begin{align}
\label{qic}
i\frac{d}{dt}
{q_i}
=\sum_{\substack{k_{j},m_{j}=0\\ j=1,2,...,l}}{c^{(i)}_{k_{1},m_{1},..,k_{l},m_{l}}}(q_1^{*})^{k_{1}}(q_1)^{m_{1}}\cdots(q_l^{*})^{k_{l}}(q_l)^{m_{l}}
(=f_i(\bf q)),
\end{align}
with $c^{(i)}_{k_{1},m_{1},..,k_{l},m_{l}}:=\lim_{N\to\infty}c^{(i,N)}_{k_{1},m_{1},..,k_{l},m_{l}}$.
Comparing this with Eq.~($\rm\ref{nonlini}$), one finds that $f_i({\bf q})$ is given by Eq.~($\rm\ref{fiqq}$), proving the proposition.

In the following, we outline the proof of Eq.~($\rm\ref{qic}$), by an argument analogous to that employed in the collective spin and spin–boson models~[49,~50].
We find it convenient to express the
$n$-point correlation function
that appears on the right-hand side of Eq.~\eqref{a61} in terms of cumulants, which allows us to extract important statistical properties order by order.
For example, the first- and second-order cumulants are given by
\begin{align}
K\bigl(O_{1}\bigr)
   \;&=\;
       \langle O_{1}\rangle,\nonumber\\
K\bigl(O_{1},O_{2}\bigr)
   \;&=\;
       \langle O_{1}O_{2}\rangle -\langle O_{1}\rangle\langle O_{2}\rangle,
\label{cumu}
\end{align}
which are nothing but the statistical average and the variance, respectively, where each $O_i$ is an annihilation or creation operator chosen from $\{\,b_1,b_1^{\dagger},\dots ,b_l,b_l^{\dagger}\,\}$.
One can express the one-and two-point correlation functions by inverting the relation Eq.~\eqref{cumu} as,
\begin{align}
\langle O_{1}\rangle
   \;&=\;
       K\bigl(O_{1}\bigr)
        ,\nonumber\\
    \langle O_{1}O_{2}\rangle      \;&=\;
    K\bigl(O_{1}\bigr)
    K\bigl(O_{2}\bigr)
    +    K\bigl(O_{1},O_{2}\bigr).
\label{cum_inverse}
\end{align}
This can be generalized to general $n$-point correlation in terms of cumulants of $m$-th order ($m\leq n$) as,
\begin{align}
\braket{O_1O_2...O_n}=\sum_{\pi\in\mathcal P_{n}}
     \prod_{B\in\pi}
K(\{ O_{j}\}_{j\in B}),
\label{corre}
\end{align}
and the $n$-th-order cumulant is defined by
\begin{equation}
K\bigl(O_{1},\dots ,O_{n}\bigr)
   \;=\;
   \sum_{\pi\in\mathcal P_{n}}
     (|\pi|-1)!\,(-1)^{|\pi|-1}
     \prod_{B\in\pi}
       \Bigl\langle\,\prod_{j\in B} O_{j}\Bigr\rangle.
\label{cumudef}
\end{equation}
Here,
$\mathcal P_{n}$ is the set of all partitions of the index set
$\{1,2,\dots ,n\}$, $|\pi|$ denotes the number of blocks in the partition $\pi\in\mathcal P_{n}$, and $B$ runs over those blocks.
The product $\prod_{j\in B} O_j$ is taken in ascending order of indices, i.e., if $B = \{j_1, j_2, \dots, j_m\}$ with $j_1 < j_2 < \cdots < j_m$, then $\prod_{j \in B} O_j := O_{j_1} O_{j_2} \cdots O_{j_m}$.

An important observation is that, if the cumulants of second or higher order vanish, i.e.,
\begin{align}
\label{alltime}
K\bigl(O_{1},\dots ,O_{n}\bigr)=0
\end{align}
for $n\geq2$, the $n$-th order correlation function is simply given by the product of the first-order cumulant (i.e., the meanfield value),
\begin{align}
\braket{O_1O_2...O_n}=\prod_{i=1}^nK\bigl(O_{i}\bigr)
=\prod_{i=1}^n \langle O_i\rangle,
\label{SIeq: n-point correlation product}
\end{align}
because the only partition which survives in the summation in the right-hand side of \eqref{corre} is $\{\{1\}, \{2\},\ldots,\{n\}\}$.
\textit{If} Eq.~\eqref{alltime} is satisfied at all times, applying Eq.~\eqref{SIeq: n-point correlation product} to Eq.~\eqref{a61}, we obtain the desired Eq.~\eqref{qic}.

We show below that Eq.~\eqref{alltime} indeed holds for our system in the large-$N$ limit under the assumption that the initial state is a tensor product of coherent states: when the state is initially given in such a form,
it will remain so at an arbitrary time.
This is shown by observing that the \textit{time derivative} of the second-order and higher cumulants vanishes [$\partial_t K(O_1, ..., O_m)=0~(m\ge2)$] whenever the $m\ge2$-order cumulants all vanish [$K(O_1, ..., O_m)=0~(m\ge2)$] at a given time, say $t=t_0$, in the large $N$ limit ($N\rightarrow \infty$), which is true when the density operator is a tensor product of coherent states~\cite{opticsbooks}.
Noting that a density operator $\rho$ can be reconstructed from the cumulants of all orders (and therefore the cumulants $K(O_1, O_2,...,O_m)$ fully characterizes the dynamics), this implies that, if the initial state satisfies $K(O_1, ..., O_m)=0~(m\ge2)$, then $K(O_1, ..., O_m)=0~(m\ge2)$ at arbitrary times ($t\ge t_0$).

For example, the time derivative of second-order cumulants is shown to vanish as follows.
Using the GKSL equation ($\rm\ref{gkslge2}$), the time derivative of second-order cumulants is given by
\begin{align}
\partial_tK\bigl(O_{1},O_{2}\bigr)
   \;&=\;
       \partial_t\langle O_{1}O_{2}\rangle -(\partial_t\langle O_{1}\rangle)\langle O_{2}\rangle-\langle O_{1}\rangle (\partial_t\langle O_{2}\rangle)\nonumber\\
&=N\Bigl(\braket{-i[O_1O_2,h]+\sum_{\mu}([l_{\mu}^{\dagger},O_1O_2]l_{\mu}-l_{\mu}^{\dagger}[l_{\mu},O_1O_2])}\nonumber\\
&-\braket{-i[O_1,h]+\sum_{\mu}([l_{\mu}^{\dagger},O_1]l_{\mu}-l_{\mu}^{\dagger}[l_{\mu},O_1])}\braket{O_2}-\braket{O_1}\braket{-i[O_2,h]+\sum_{\mu}([l_{\mu}^{\dagger},O_2]l_{\mu}-l_{\mu}^{\dagger}[l_{\mu},O_2])}\Bigr).
\label{Kdown}
\end{align}
Let us first concentrate on the terms involving the Hamiltonian $H=Nh$:
\begin{align}
\partial_tK_H\bigl(O_{1},O_{2}\bigr)
:=
\braket{i[O_1,Nh](\braket{O_2}-O_2)}+\braket{(\braket{O_1}-O_1)i[O_2,Nh]}
.
\label{K2}
\end{align}
We find it useful to apply normal ordering (denoted by the symbol $:\ :$) as
\begin{align}
\partial_tK_H\bigl(O_{1},O_{2}\bigr)=
\braket{:i[O_1,Nh](\braket{O_2}-O_2):}+\braket{:(\braket{O_1}-O_1)i[O_2,Nh]:}
+\mathcal{O}\Bigl(\frac{1}{N}\Bigr).
\label{K2order}
\end{align}
Here, we used the fact that the commutation relation $[b_i,b_j^\dagger]=\delta_{i,j}/N$ (that appears in the normal ordering process) is at most of $\mathcal{O}(1/N)$.

Under the assumption that the initial state is given by a tensor product of coherent states (i.e., the second or higher cumulants vanish at a given time $t=t_0$), the creation and annihilation operators in the first and the second terms on the left-hand side are replaced with complex numbers (thanks to normal ordering), allowing the replacement $O_{i}\rightarrow \langle O_{i}\rangle, i=1,2$. Hence, these two terms vanish, and as a result,
the right-hand side of Eq.~($\rm\ref{K2order}$) vanishes in the large-$N$ limit,
\begin{align}
\lim_{N\to\infty}\partial_t K_H\bigl(O_{1},O_{2}\bigr)
|_{t=t_0}
   \;&=\; 0.
\end{align}
Similarly, the dissipative part of the right-hand side of \eqref{Kdown}
\begin{align}
    \partial_t K_L := \sum_{\mu}\Bigl(\braket{([l_{\mu}^{\dagger},O_1]l_{\mu}-l_{\mu}^{\dagger}[l_{\mu},O_1])(O_2-\braket{O_2}}-\braket{(O_1-\braket{O_1})([l_{\mu}^{\dagger},O_2]l_{\mu}-l_{\mu}^{\dagger}[l_{\mu},O_2])}\Bigr)
\end{align}
can be shown to vanish at $t=t_0$ with the same condition stated above in the large-$N$ limit:
\begin{align}
    \lim_{N\rightarrow \infty}\partial_t K_L
    |_{t=t_0}
    = 0.
\end{align}
As a result,
one obtains
\begin{align}
\lim_{N\to\infty}\partial_t K\bigl(O_{1},O_{2}\bigr)
  |_{t=t_0}
   \;&=\;
   0.
\end{align}

More generally, in a similar manner to the above, one can show that, if the time derivative of the $(n-1)$-th-order cumulants vanishes at the initial time $t=t_0$, then the derivative of the $n$-th-order cumulants also vanishes
in the large-$N$ limit:
\begin{align}
\lim_{N\to\infty}\partial_tK\bigl(O_{1},O_{2},...,O_n\bigr)|_{t=t_0}
   \;&=\; 0.
\end{align}
This proves that the dynamics is fully characterized by the meanfield dynamics
and hence proves the desired result~\eqref{qic}, and hence the preposition.
\end{proof}

\endgroup

\renewcommand{\theequation}{C.\arabic{equation} }
\setcounter{equation}{0}
\section{C. Linear stability analysis for bipartite bosonic systems with L-$\mathcal{PT}$ symmetry and conserved particle number}
\renewcommand{\thefigure}{C.\arabic{figure} }
\setcounter{figure}{0}
We now focus on a bipartite bosonic system with a uniform solution. Denote the two sublattices by
$A$ and
$B$, the uniform solution is characterized by two order parameters, $q_A:=2 \sum_{i\in A}q_i/l$ and $q_B:=2\sum_{i\in B}q_i/l$. Introducing the vector ${\bf{q}}:=(q_{A},q_{B})^{T}$, the mean-field equation takes the compact form $i\partial_t{\bf{q}}={\bf{f}}({\bf{q}})$, ${\bf{f}}({\bf{q}}):=(f({\bf{q}}),f({\tilde{P}\bf{q}^{*}}))^{T}$, where $\tilde P$ exchanges the two components and
$f:\mathbb{C}^2\to\mathbb{C}$.

Writing order parameters on each sublattice in polar form,
$q_{A(B)}=r_{A(B)}e^{i\theta_{A(B)}}$, and defining the phase difference $\Delta\theta:=\theta_A-\theta_B$, we obtain:
\begin{align}
\label{ap}
i\partial_{t}r_{A}-r_A \partial_{t}\theta_A&=g(r_A,r_B,i\Delta\theta),\nonumber\\
i\partial_{t}r_{B}-r_B \partial_{t}\theta_B&=g^*(r_B,r_A,i\Delta\theta),
\end{align}
with a complex function  $g:\mathbb{C}^3\to\mathbb{C}$. Separating real and imaginary parts, $g:=g_{\rm{re}}+ig_{\rm{im}}$, and replacing $i\Delta\theta\to\Delta\theta$, Eq.~($\rm\ref{ap}$) becomes
\begin{align}
\partial_{t}r_{A}&=g_{\rm{im}}(r_A,r_B,\Delta\theta)\nonumber\\
\partial_{t}r_{B}&=-g_{\rm{im}}(r_B,r_A,\Delta\theta)\nonumber\\
\partial_{t}\Delta\theta&=-\frac{1}{r_A}g_{\rm{re}}(r_A,r_B,\Delta\theta)+\frac{1}{r_B}g_{\rm{re}}(r_B,r_A,\Delta\theta).
\end{align}

{Using particle-number conservation, $r_A^2+r_B^2=\textit{const}$, it is convenient to introduce}
\begin{align}
G(r_A,r_B,\Delta\theta):=r_Ag_{\rm{im}}(r_A,r_B,\Delta\theta)=r_Bg_{\rm{im}}(r_B,r_A,\Delta\theta),
\end{align}
which is symmetric under the exchange $r_A\leftrightarrow r_B$, and
\begin{align}
H(r_A,r_B,\Delta\theta):=-\frac{1}{r_A}g_{\rm{re}}(r_A,r_B,\Delta\theta)+\frac{1}{r_B}g_{\rm{re}}(r_B,r_A,\Delta\theta),
\end{align}
which is antisymmetric, $H(r_A,r_B,\Delta\theta)=-H(r_B,r_A,\Delta\theta)$. {Expressed in terms of the pair ($G$,$H$), the mean-field dynamics reduces to the set of equations reported as Eq.~\eqref{rarbthe} in the main text.}

Subsequently, we perform the linear stability analysis for Eq.~($\ref{rarbthe}$). The Jacobian matrix around a fixed point takes the form
\begin{align}
\label{Jge}
J=\left.\begin{pmatrix}
\gamma_{1}&\alpha\\
\beta&\gamma_{2} \\
\end{pmatrix}\right|_{ss}:=\left.\begin{pmatrix}
-\frac{1}{r_{B}}\frac{\partial G}{\partial r_{B}}+\frac{1}{r_{A}}\frac{\partial G}{\partial r_{A}}&-\frac{1}{r_{B}}\frac{\partial G}{\partial \Delta\theta}\\
\frac{\partial H}{\partial r_{B}}-\frac{r_{B}}{r_{A}}\frac{\partial H}{\partial r_{A}}&\frac{\partial H}{\partial \Delta\theta} \\
\end{pmatrix}\right|_{ss},
\end{align}
where we eliminated $r_{A}$ degrees of freedom since the particle number is conserved, $r_{A}^{2}+r^{2}_{B}=\textit{const.}$

If the fixed point is PT-symmetric, namely $r_{A}=r_{B}$, the Jacobian simplifies to
\begin{align}
J=\left.\begin{pmatrix}
0&\alpha\\
\beta&0 \\
\end{pmatrix}\right|_{ss}.
\end{align}
The collective excitation spectrum is given by
$\lambda=\pm\sqrt{\alpha\beta}$ with corresponding eigenmodes: $\delta{\bf{q}}:=(\delta r_{B},\delta\Delta\theta)=(\pm\sqrt{\alpha/\beta},1)^T$. Thus, in the physical case, the fixed point is a center, indicating persistent oscillations.

In contrast, for the PT-broken fixed points ($r_{A}\neq r_{B}$), the collective excitation spectrum is given by
\begin{align}
\lambda= R\pm\sqrt{R^{2}+Q},
\end{align}
with $R:=(\gamma_{1}+\gamma_{2})/2$ and $Q:=(\alpha\beta-\gamma_{1}\gamma_{2})/4$. Because $G$ ($H$) is symmetric (anti-symmetric) under $r_A\leftrightarrow r_B$, the coefficients $\gamma_{1}$ and $\gamma_{2}$ in Eq.~($\rm\ref{Jge}$) are antisymmetric, while the product $\alpha\beta$ is symmetric.
Under the assumption that $R\neq0$, a pair of PT-broken solutions is \textit{physically} admissible only if $\gamma_1\gamma_2>\alpha\beta$; in that case, one fixed point of the pair is stable and the other unstable. When $\gamma_1\gamma_2<\alpha\beta$, at least one eigenvalue always has a positive real part for both solutions, rendering them intrinsically unstable.
Note that a CEP typically emerges from the ordered phase to the continuous phase transition point with $r_A=r_B$ and $\alpha\beta=0$.

\renewcommand{\theequation}{D.\arabic{equation}}
\setcounter{equation}{0}
\section{D. Analysis of L-$\mathcal{PT}$-symmetric models}
\renewcommand{\thefigure}{D.\arabic{figure}}
\setcounter{figure}{0}

Our symmetry-based framework can be applied to a wide range of classes. This includes one-collective spin models, which have been studied thoroughly in the context of DCTCs, and spatially extended bipartite bosonic systems that preserve particle number. In Section D.1, we deal with a general one-collective spin model. In Sections D.2, D.3, we further analyze the generalized DDM and the dissipative LMG model, respectively.
In Section D.4, we deal with a one-collective spin model without $\mathbb{Z}_2$ symmetry of the Hamiltonian.
Finaly, in Section D.5, we analyze a spatially extended bipartite many-body bosonic system with conserved particle number.

\subsection{D.1 General dissipative one-collective spin model}
There are (infinitely in theory) many one-collective spin models to which our theory can be applied.
\textcolor{black}{For concreteness, we apply our framework to models described by the following L-$\mathcal{PT}$-symmetric GKSL equation:
\begin{align}
\partial_{t}\rho=-i[H_{\rm PT},\rho]+\frac{1}{2}\sum_{\mu}(\hat{\mathcal{D}}[L_{\mu}]+\hat{\mathcal{D}}[PTL_{\mu}^\dagger (PT)^{-1}])\rho,
\end{align}
where we defined $H_{\rm PT}:=(H+PTH(PT)^{-1})/2$. When we choose $PT=\prod_i\sigma_x^i K$, the general PT-symmetric Hamiltonian and Lindblad operators can be written as
\begin{eqnarray}
H_{\rm PT}=S\sum_{l,m,n}g_{l,m,n}(m_x)^l(m_y)^m(m_z)^{2n},
\end{eqnarray}
and
\begin{eqnarray}
L_\mu=\sqrt{S}\sum_{l,m,n}\gamma_{l,m,n,\mu}(m_x)^l(m_y)^m(m_z)^{n},
\end{eqnarray}
respectively.}
Here $m_\alpha$ ($\alpha=x,y,z$) is the normalized collective spin operator, and $S$ is the conserved total spin, and the constant $g_{l,m,n}$ ($\gamma_{l,m,n,\mu}$) is a real number (complex) with $n,m,l\in\mathbb{N}_0$.
Parity and time-reversal operators act on normalized collective spin operators as
\begin{align*}
P:\ \ m_x\to m_x,\ \ m_y\to -m_y,\ \  m_z\to -m_z\\
T:\ \ m_x\to m_x,\ \ m_y\to -m_y,\ \  m_z\to m_z,
\end{align*}
respectively.

Importantly, our theory suggests that the DCTC is extremely robust against a wide range of unwanted dissipation processes, since dissipators that respect L-$\mathcal{PT}$ symmetry include representative dissipation processes such as pump $L=m_+$, decay $L=m_-$ and dephasing $L=m_x,m_y,m_z$ (i.e. $\hat{\mathcal{D}}[L]=\hat{\mathcal{D}}[PTL^\dagger (PT)^{-1}]$).

\subsection{D.2 Continuous L-$\mathcal{PT}$ phase transition  --- Generalized DDM}
We first consider the generalized DDM with the Hamiltonian $H=S(2gm_{x}+\omega m_z^2)$ and the Lindblad operator $L=\sqrt{\kappa S}m_{-}$.
The mean-field equation is given by Eq.~($\ref{timematrix1}$) in the main text.
In this model, there are four distinct fixed points; two PT-symmetric ones ${\bf{M}}_{\pm,\rm{PT}}$ and two PT-broken ones ${\bf{M}}_{\pm,\rm{PTb}}$. Former and latter solutions are given by
\begin{eqnarray}
\label{fixedPT}
{\bf{M}}_{\pm,\rm{PT}}=\left(\pm\sqrt{1-\left(\frac{\kappa}{g}\right)^2},\frac{\kappa}{g},0\right),
\end{eqnarray}
and
\begin{eqnarray}
\label{fixedPTb}
{\bf{M}}_{\pm,\rm{PTb}}=\left(\frac{g\omega}{\kappa^2+\omega^2},\frac{g\kappa}{\kappa^2+\omega^2},\pm\sqrt{1-\frac{g^2}{\kappa^2+\omega^2}}\right),
\end{eqnarray}
respectively.

\begin{figure}[hbtp]
   \vspace*{-3.5cm}
    \hspace*{-11.3cm}
\includegraphics[bb=0mm 0mm 90mm 150mm,width=0.26\linewidth]{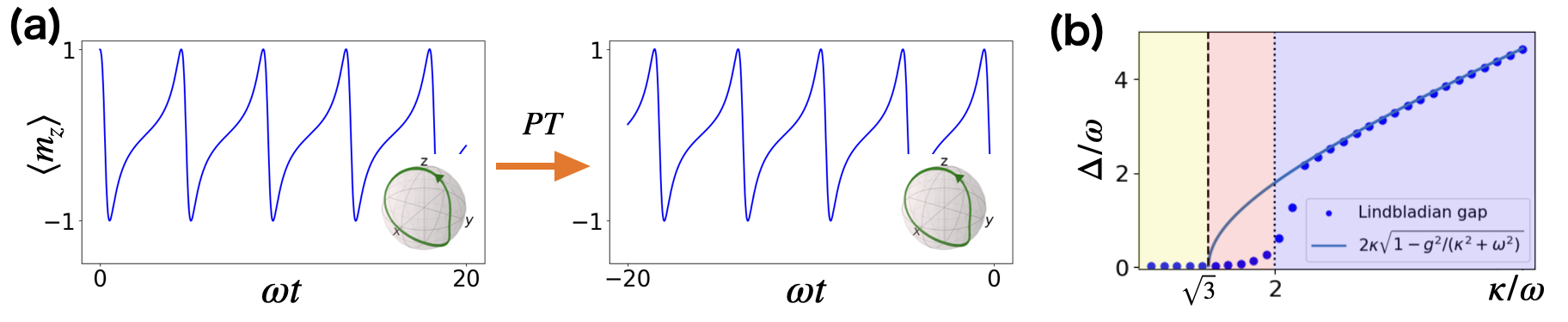}
    \caption{\small{\justifying(a) Dynamics of magnetization and the PT transformed one ($\braket{m_{z}}\to-\braket{m_{z}},\ t\to-t$) with $g=2$, $\omega=1$, and $\kappa=1.7$. inset: Their corresponding trajectory of magnetization's dynamics. They describe the same attractors. (b) The Lindbladian gap $\Delta_{\rm{gap}}$ for $S=200$ (blue dot) and the mean-field value $\Delta_{\rm{mean}}$ (light blue line, where $\kappa_{1c}/\omega=\sqrt{3}$,\ $\kappa_{2c}/\omega=2$}}
    \label{timeevo}
\end{figure}

Let us perform the linear stability analysis for the $Y-Z$ plane.
Excitation spectra and modes at the PT-symmetric solutions ${\bf{M}}_{\pm,\rm{PT}}$ $\eqref{fixedPT}$ are given by $\lambda=\{\sqrt{\alpha\beta},-\sqrt{\alpha\beta}\}$ and $\delta {\bf{M}}=\{(\sqrt{\alpha/\beta},1)^T,\ (-\sqrt{\alpha/\beta},1)^T\}$ with
\begin{align}
\sqrt{\alpha\beta}&=2i(g^2-\kappa^2)^{1/4}(\sqrt{g^2-\kappa^2}\mp \omega)^{1/2},\\
\sqrt{\alpha/\beta}&=i(g^2-\kappa^2)^{1/4}(\sqrt{g^2-\kappa^2}\mp\omega)^{1/2}/g,
\end{align}
where $\delta {\bf{M}}:=(\delta Y,\delta Z)$. The solution ${\bf{M}}_{+,\rm{PT}}$ is a center for $\kappa<\sqrt{g^2-\omega^2}$, and the solution ${\bf{M}}_{-,\rm{PT}}$ is also a center for $\kappa<g$.
These transition points $\kappa=g$ and $\kappa=\sqrt{g^2-\omega^2}$ are CEPs.

While, for the PT-broken solutions ${\bf{M}}_{\pm,\rm{PTb}}$ $\eqref{fixedPT}$, the excitation spectra and modes are given by
\begin{align}
\lambda&=\{2(\kappa+ i\omega){M}_{\pm,\rm{PTb},z},\ 2(\kappa- i\omega){M}_{\pm,\rm{PTb},z}\},\\
\delta {\bf{M}}&=\{((\kappa+i\omega){M}_{\pm,\rm{PTb},z}/g,1)^T,\ ((\kappa-i\omega){M}_{\pm,\rm{PTb},z}/g,1)^T\},
\end{align}
where, ${M}_{\pm,\rm{PTb},z}$ is the $z$-component of ${\bf{M}}_{\pm,\rm{PTb}}$. The solution ${\bf{M}}_{+,\rm{PTb}}$ is unstable for any $\kappa$, and the solution ${\bf{M}}_{-,\rm{PTb}}$ is stable for $\kappa>\sqrt{g^2-\omega^2}$. Moreover, a CEP also appears in $\kappa=\sqrt{g^2-\omega^2}$ from the PPTB phase.

So far, we have classified the PT symmetry for fixed points. Now, we explore the PT symmetry of time-dependent solutions (e.g. oscillating solution) as well. To proceed, we first define the PT symmetry for time-dependent solutions.
We say that the PT symmetry of a time-dependent solution
${\bf{q}}(t)$ is unbroken if ${\bf{q}}(t)$
is equivalent to
$\tilde{P}\tilde{T}{\bf{q}}(t)=\tilde{P}{\bf{q}^*}(-t)$
at late times (i.e. the attractor),
while, if not, it is spontaneously broken.
Here, we say that the two solutions are equivalent when they can be transformed from one to the other using symmetry operations present in the system (other than the n-PT symmetry).
For example, a mean-field equation of a system governed by a time-independent Lindbladian always has continuous time-translation symmetry.
In this case, we say that ${\bf{q}}(t)$ is PT-symmetric if ${\bf{q}}(t)=\tilde{P}{\bf{q}^*}(-t+t_{0})$, where $t_{0}\in \mathbb{R}$.

Fig.$\text{\ref{timeevo}}$ (a) shows magnetization dynamics and its PT-transformed one in the PT phase. As expected in the presence of a center, persistent oscillations exist. Moreover, these oscillations are PT-symmetric since the trajectory is the same as the PT transformed one, that is, ${\bf{q}}(t)=\tilde{P}{\bf{q}^*}(-t+t_{0})$ holds.

Next, we compare the value of the gap of the matrix $J$ (\ref{LL}) in the linear stability analysis to the Lindbladian gap, which is the absolute value of the real part of the second maximal eigenvalue $\Delta_{\text{gap}}:=|\max_{i\neq0}\textrm{Re}[\lambda_{i}]|$ (Fig.~$\rm\ref{timeevo}$ (b)). One can expect that the Lindbladian gap vanishes in the PT and PPTB phase because of the presence of persistent oscillations without decay, while in the FPTB phase, it corresponds to the real part of mean-field excitation spectra ${\rm{Re}}[\lambda]=2\kappa\left.Z\right|_{ss}$ with $\left.
Z\right|_{ss}=-\sqrt{(\kappa^2-\kappa_{1c}^2)/(\kappa^2+\omega^2)}$.
Fig.~$\ref{lpt}$ (d) indicates that the mean-field prediction coincides well with the Lindbladian gap for a large $S$, as expected.

\begin{figure}[hbpt]
   \vspace*{-2.0cm}
     \hspace*{-5cm}
\includegraphics[bb=0mm 0mm 90mm 150mm,width=0.23\linewidth]{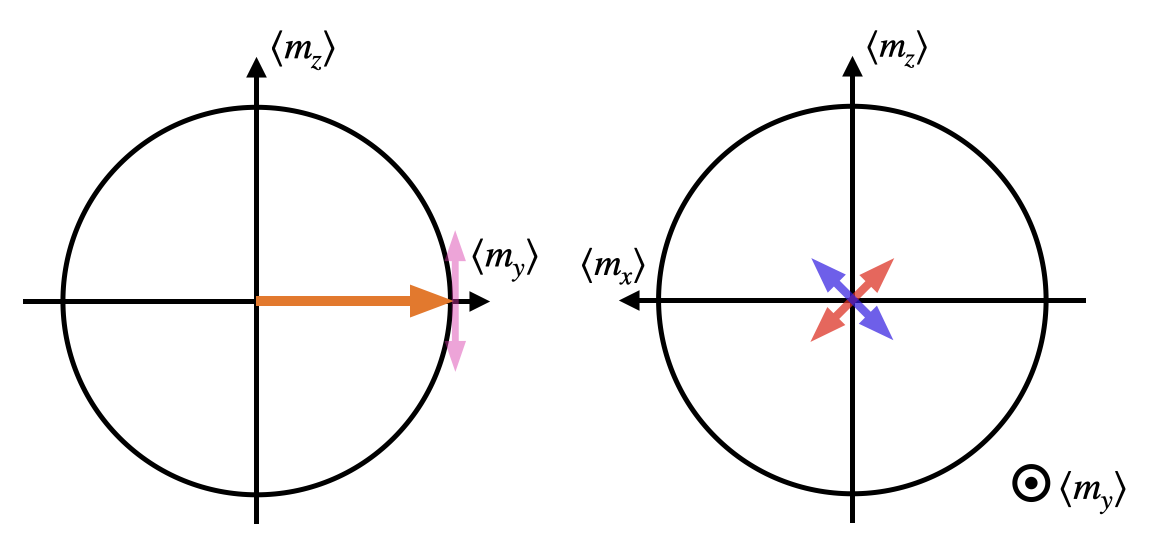}
    \caption{\small{\justifying Left: The collective excitation mode (pink arrow) and the magnetization's vector of the fixed point (orange arrow) at the transition point $\kappa=g$ for the DDM. Right: The collective excitation modes (blue and red or pink arrows) at the transition point. These are orthogonal. Here, the components of collective excitation modes are generally complex numbers, so we take absolute values for complex elements.}}
    \label{xyz}
\end{figure}

\textcolor{black}{Finally, we comment on the special case $\omega=0$, which corresponds to the DDM. In this scenario, the effects of CEPs cannot be observed. To see why, recall that revealing a CEP requires an initial perturbation that departs from the coalescing eigenmode $\delta {\bf{M}}=(0,1)$; one must excite a component with
$\delta {\bf{M}}=(a,b)$ with $a\neq0,\ a,b\in\mathbb{C}$. }
\textcolor{black}{In the DDM, however, the collective magnetization is restricted to the plane $(Y,Z)=(1,0)$, which is orthogonal to
$\delta {\bf{M}}=(0,1)$ (as illustrated in the left diagram of Fig.$\text{\ref{xyz}}$). The system is therefore unable to evolve in the $Y$-direction, and the CEP-related divergence of dynamical time-scales is kinematically blocked.}

\textcolor{black}{One might attempt to excite the system along the $X$-direction instead, but a direct calculation shows that, in the $X-Z$ plane, the Jacobian at the PT-symmetric fixed point has the same structure as Eq.~$\eqref{LL}$, but $\alpha$ and $\beta$ simultaneously vanish at the transition point, indicating the absence of a CEP.
Hence, the DDM with $\omega=0$ provides an instructive example in which the L-$\mathcal{PT}$ symmetry is present, yet the CEP effects are absent for purely geometric reasons.}

\subsection{D.3 Discontinuous L-$\mathcal{PT}$ phase transition --- Dissipative LMG model.}
\begin{figure}[hbpt]
   \vspace*{-1.9cm}
     \hspace*{-4.3cm}
\includegraphics[bb=0mm 0mm 90mm 150mm,width=0.25\linewidth]{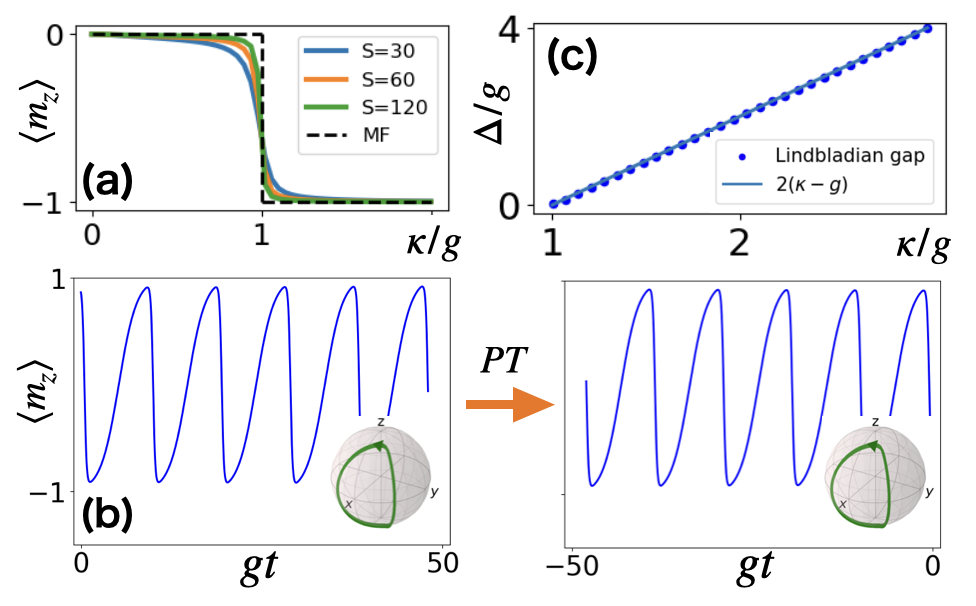}
    \caption{ \small{\justifying The numerical analysis of the dissipative LMG model. (a) The normalized magnetization $\braket{m_{z}}$ in the stationary state $\lim_{t\to\infty}\rho(t)$ for finite $S$ and mean field solution ($S=\infty$). (b) A time evolution of magnetization and the PT transform ($\braket{m_{z}}\to-\braket{m_{z}},\ t\to-t$) at $\kappa/g=0.8$. inset: Their corresponding trajectories. They describe the same attractor. (c) The Lindbladian gap $\Delta_{\text{gap}}$ for $S=30$ (blue dot) and the mean-field value $\Delta_{\text{mean}}$=2$(\kappa-g)$ (light blue line).} }
    \label{dlpt}
\end{figure}

We next analyze the Lipkin-Meshkov-Glick (LMG) Hamiltonian $H=gS(m_{+}^{2}+m_{-}^{2})/2S$ with collective decay $L=\sqrt{\kappa S}m_{-}$.
In this model, a discontinuous phase transition occurs at $\kappa=g$ as shown in Fig.$\rm\ref{dlpt}$ (a). This model has an L-$\mathcal{PT}$ symmetry \eqref{HuberPT}.

The time evolution in the large $S$ limit is written down as
\begin{align}
\label{D6}
\partial_{t}{\bf{M}}
= 2\begin{pmatrix}
-gYZ+\kappa XZ\\
-gXZ+\kappa YZ\\
2gXY-\kappa(1-Z^2)\\
\end{pmatrix}.
\end{align}
The nonlinear dynamical system ($\rm\ref{D6}$) also has the n-PT symmetry ($\ref{nonlinearPT}$) with $\tilde{P}=diag(1,1,-1)$.

There are six distinct fixed points: four PT-symmetric and two PT-broken ones. The PT-symmetric solutions are given by
\begin{eqnarray}
\label{fixed}
{\bf{M}}_{\pm,\rm{PT}}^{\text{I}}=\left(M_\pm,M_\mp,0 \right),\ \ \ {\bf{M}}_{\pm,\rm{PT}}^{\text{I\hspace{-1.2pt}I}}=\left(-M_\pm,-M_\mp,0 \right),
\end{eqnarray}
with $M_\pm:=\sqrt{\frac{1\pm\sqrt{1-(\kappa/g)^2}}{2}}$.
On the other hand, the PT-broken solutions are given by
\begin{eqnarray}
{\bf{M}}_{\pm,\rm{PTb}}=\left(0,0,\pm1 \right).
\end{eqnarray}

The elements in the Jacobian $\eqref{LL}$ at PT-symmetric fixed points are given by
\begin{align}
\alpha=-2gX+2\kappa Y,\ \ \beta=\frac{4}{X}(X^2-Y^2),
\end{align}
indicating that they are centers for $\kappa<g$, indicating the presence of persistent oscillations. This is consistent with Ref.\cite{Lee} that reported that
a periodic solution emerges, as shown in Fig.$\rm\ref{dlpt}$ (b).
Moreover, by performing a PT transformation ($\braket{m_{z}}\to-\braket{m_{z}}$, $t\to -t$), one finds that the system describes the same attractor, confirming that the solution is PT-symmetric. Additionally, this analysis shows that at the transition point $\kappa=g$, both coefficients vanish $\alpha=\beta=0$. As a result, the oscillation period diverges at this transition, but this divergence does not correspond to a CEP.

The excitation spectra and modes around the $\mathcal{PT}$-broken fixed points ${\bf{M}}_{\pm,\rm{PTb}}$ are given by
\begin{align}
\lambda&=\{2(\kappa+ g){{M}}_{\pm,\rm{PTb,z}},\ 2(\kappa- g){{M}}_{\pm,\rm{PTb,z}}\},\\
\delta {\bf{M}}&=\{(1,-1)^T,\ (1,1)^T\}.
\end{align}
These expressions indicate that the solution ${\bf{M}}_{+,\rm{PTb}}$ is unstable for any $\kappa$, whereas the solution ${\bf{M}}_{-,\rm{PTb}}$ is stable for $\kappa<g$. At the transition point $\kappa=g$, the solution ${\bf{M}}_{-,\rm{PTb}}$ becomes destabilized, but notably, a CEP does not emerge from the PT-broken phase as well.
Fig.$\rm\ref{dlpt}$ (c) shows the Lindbladian gap and the mean-field excitation spectrum. These results demonstrate that the mean-field prediction accurately captures the discontinuous L-$\mathcal{PT}$ phase transition in the limit of large $S$.

\subsection{D.4  One-collective spin model without $\mathbb{Z}_2$ Hamiltonian}
We consider a
light-matter system composed of emitters coupled to light
modes which propagate in a waveguide, where the Hamiltonian and Lindblad operators are given by
\begin{align}
\label{emi}
H=S(2g m_x-\omega\gamma\{m_x,m_y\})
\end{align}
and
\begin{align}
L_1=\sqrt{\gamma S/2}((2\omega+1)m_x-im_y),\ \ \ L_2=\sqrt{\gamma S/2}\ m_-,
\end{align}
respectively.
The parameters $\omega, g, \gamma$ denote the Rabi frequency, feedback strength, and decay rate. This model exhibits a DCTC as shown in \cite{Buonaiuto}.

Contrary to the expectation in Ref.\cite{Piccitto}, the Hamiltonian does not have $\mathbb{Z}_2$ symmetry since there are no spin unitary operators that invert only one component ($m_\alpha\to -m_\alpha$, $\alpha=x,y,z$).
On the other hand, the Lindbladian has the L-$\mathcal{PT}$ symmetry with $P=(i)^{2S}\exp(i\pi S m_x)$ and $T=K$.
The time evolution in the large $S$ limit is written down as
\begin{align}
\label{D8}
\partial_{t}{\bf{M}}
= 2\begin{pmatrix}
\gamma XZ\\
-gZ+\gamma\kappa YZ\\
gY-\gamma X^2-\gamma\kappa Y^{2}\\
\end{pmatrix},
\end{align}
with $\kappa=2\omega+1$.
As expected, the nonlinear dynamical system ($\rm\ref{D8}$) has n-PT symmetry ($\ref{nonlinearPT}$) with $\tilde{P}=diag(1,1,-1)$. Moreover, the Jacobian of PT-symmetric fixed points is given by
\begin{align}
J=2\left.\begin{pmatrix}
0&-g+\gamma\kappa Y\\
g-2\gamma\omega Y&0 \\
\end{pmatrix}\right|_{ss},
\end{align}
indicating that it is a center for a physical case and the transition point becomes a CEP.

Thus, this example demonstrates that the $\mathcal{PT}$ symmetry of Lindbladians plays a crucial role for emergence of DCTCs, instead of the $\mathbb{Z}_2$ symmetry of the Hamiltonian.

\section{D.5. $d$-dimensional L-$\mathcal{PT}$-symmetric boson model with local interaction}
Finally, we analyze a $\mathcal{PT}$-symmetric spatially extended bipartite bosonic system that preserves particle number. Specifically, we consider a $d$-dimensional bosonic Hamiltonian with local hopping and potential with rates $g$ and $\omega$ on a bipartite square hyperlattice
\begin{align}
\label{ham}
H=\sum_{<i,j>}g(a^{\dagger}_{i\in A}a_{j\in B}+a_{i\in A}a^{\dagger}_{j\in B})-\omega^\prime a^{\dagger}_{i\in A}a_{i\in A}a^{\dagger}_{j\in B}a_{j\in B},
\end{align}
and incoherent hopping from the $A$ sublattice to the $B$ sublattice, with a uniform rate $\kappa^\prime$,
\begin{eqnarray}
\label{lin}
\dot{\rho}=-i[H,\rho]+\sum_{<i,j>}\kappa^\prime \hat{\mathcal{D}}[a_{i\in A}a_{j\in B}^{\dagger}]\rho.
\end{eqnarray}
Here, $<i,j>$ means the set of nearest neighbor sites and the total particle number $N:=\sum_ia^\dagger_i a_i$ is conserved (i.e., the Lindbladian has (strong) $U(1)$ symmetry). This model also has an L-$\mathcal{PT}$ symmetry, where parity is the exchange of sublattices $A$ and $B$, and a time-reversal operator is a complex conjugate operator.

Below, we consider the case where the interaction and dissipation strength $\omega^\prime,\kappa^\prime$ are rescaled to $\omega=\omega^\prime/N,\ \kappa=\kappa^\prime/N$, indicating that the coherent interaction and dissipation terms are weak in the large-$N$ limit. We introduce the rescaled bosonic operators as $b_i=a_i/\sqrt{N}$.
Under this scaling, the Hamiltonian ($\rm\ref{ham}$) and the GKSL equation ($\rm\ref{lin}$) can be rewritten in the form $H=:hN$ with
\begin{align}
h=\sum_{<i,j>}g(b^{\dagger}_{i\in A}b_{j\in B}+b_{i\in A}b^{\dagger}_{j\in B})-\omega b^{\dagger}_{i\in A}b_{i\in A}b^{\dagger}_{j\in B}b_{j\in B}
\end{align}
and
\begin{eqnarray}
\dot{\rho}=N(-i[h,\rho]+\sum_{<i,j>}\kappa \hat{\mathcal{D}}[b_{i\in A}b_{j\in B}^{\dagger}]\rho).
\end{eqnarray}

Taking the large-$N$ limit, the mean-field equation is described by
\begin{align}
i\partial_tq_{i\in A}=\sum_{j\in \mathbb{O}_{i,j}}(gq_{j\in B}-i\kappa q_{i\in A}|q_{j\in B}|^{2}-\omega q_{i\in A}|q_{j\in B}|^2),\nonumber\\
i\partial_tq_{j\in B}=\sum_{i\in \mathbb{O}_{j,i}}(gq_{i\in A}+i\kappa q_{j\in B}|q_{i\in A}|^{2}-\omega q_{j\in B}|q_{i\in A}|^2),
\end{align}
with $q_{i\in A}:=\lim_{N\to\infty}\braket{b_{i\in A}},\ q_{j\in B}:=\lim_{N\to\infty}\braket{b_{j\in B}}$, where
$\mathbb{O}_{i,j}$ denote the set of the nearest sites $j$ from the site $i$.
As guaranteed by Theorem in the main text, the mean-field equation satisfies the n-PT symmetry ($\ref{nonlinearPT}$).
Furthermore, assuming a uniform solution, it reduces
\begin{align}
i\partial_t q_A=2d(g q_B-i\kappa q_A|q_B|^{2}-\omega  q_A| q_B|^2),\ \ \
i\partial_t q_B=2d(g q_A+i\kappa q_B|q_A|^{2}-\omega q_B|q_A|^2),
\end{align}
with $q_A:=2 \sum_{i\in A}q_i/l$ and $q_B:=2\sum_{i\in B}q_i/l$.
In the polar representation $q_A=r_{A}e^{i\theta_{A}}$, $q_B=r_{B}e^{i\theta_{B}}$, it can be rewritten as
\begin{align}
\label{D17}
\partial_t r_{A}&=-2d(\kappa r_{A}r_{B}^{2}+gr_{B}\sin(\Delta\theta)), \ \ \ \partial_t\Delta\theta=-2d\left(g\left(\frac{r_{B}}{r_{A}}-\frac{r_{A}}{r_{B}}\right)\cos(\Delta\theta)+\omega(r_A^2-r_B^2)\right),\nonumber\\
\partial_t r_{B}&=2d(\kappa r_{A}^{2}r_{B}+gr_{A}\sin(\Delta\theta)),
\end{align}
with $\Delta\theta:=\theta_{a}-\theta_{b}$ and $r_A^2+r_B^2=1$.
The mean-field equation ($\rm\ref{D17}$) is equivalent to the one for the generalized DDM ($\rm\ref{timematrix1}$) in the Schwinger boson representation, except for the multiple constant $2d$. Hence, continuous time-translation symmetry is spontaneously broken into a discrete one, and the transition point is associated with spontaneous n-PT symmetry breaking and becomes a CEP.


\begin{thebibliography}{10}
\makeatletter
\providecommand{\DOTSB}{}
\providecommand{\sum@}{\sum}
\providecommand{\slimits@}{}

\bibitem{Wilczek}
F. Wilczek, Quantum time crystals, Phys. Rev. Lett. $\textbf{109}$, 160401 (2012).

\bibitem{Sacha}
K. Sacha and J. Zakrzewski, Time crystals: a review, Rep. Prog. Phys. $\textbf{81}$ 016401 (2018).

\bibitem{Else}
D. V. Else, C. Monroe, C. Nayak, and N. Y. Yao, Discrete time crystals, Annu. Rev. Condens. Matter Phys. $\textbf{11}$, 467 (2020).

\bibitem{Watanabe}
H. Watanabe and M. Oshikawa, Absence of quantum time crystals, Phys. Rev. Lett. $\textbf{114}$, 251603 (2015).

\bibitem{Iemini}
F. Iemini, A. Russomanno, J. Keeling, M. Schir\`{o}, M. Dalmonte, and R. Fazio, Boundary time crystals, Phys. Rev. Lett. $\textbf{121}$, 035301 (2018).

\bibitem{Piccitto}
G. Piccitto, M. Wauters, F. Nori, and N. Shammah, Symmetries and conserved quantities of boundary time crystals in generalized spin models, Phys. Rev. B $\textbf{104}$, 014307 (2021).

\bibitem{dos}
L. F. dos Prazeres, L. da S. Souza, and F. Iemini, Boundary time crystals in collective $d$-level systems, Phys. Rev. B $\textbf{103}$, 184308 (2021).

\bibitem{Buonaiuto}
G. Buonaiuto, F. Carollo, B. Olmos, and I. Lesanovsky, Dynamical phases and quantum correlations in an emitter-waveguide system with feedback, Phys. Rev. Lett. $\textbf{127}$, 133601 (2021).

\bibitem{Minganti4}
F. Minganti, I. I. Arkhipov, A. Miranowicz, and F. Nori, Correspondence between dissipative phase transitions of light and time crystals, arXiv:2008.08075 [quant-ph] (2020).

\bibitem{Lled2}
C. Lled\'{o} and M. H. Szyma\'{n}ska, A Dissipative time crystal with or without Z$_2$ symmetry breaking, New J. Phys. $\textbf{22}$ 075002 (2020).

\bibitem{Cabot}
A. Cabot, G. L. Giorgi, and R. Zambrini, Nonequilibrium Transition between Dissipative Time Crystals, PRX Quantum $\textbf{5}$, 030325 (2024).

\bibitem{Li}
Y. Li, C. Wang, Y. Tang, and Y.-C. Liu, Time Crystal in a Single-Mode Nonlinear Cavity, Phys. Rev. Lett. $\textbf{132}$, 183803 (2024).

\bibitem{Booker}
C. Booker, B. Bu\v{c}a, and D. Jaksch, Non-stationarity and dissipative time crystals: spectral properties and finite-size effects, New J. Phys. $\textbf{22}$ 085007 (2020).

\bibitem{Passarelli}
G. Passarelli, P. Lucignano, R. Fazio, and A. Russomanno, Dissipative time crystals with long-range Lindbladians, Phys. Rev. B $\textbf{106}$, 224308 (2022).

\bibitem{Ya-Xin}
Ya-Xin Xiang, Qun-Li Lei, Zhengyang Bai, Yu-Qiang Ma, Self-organized time crystal in driven-dissipative quantum system, Phys. Rev. Research $\textbf{6}$, 033185 (2024).

\bibitem{Yang}
S. Yang, Z. Wang, L. Fu, and J. Jie, Emergent continuous time crystal in dissipative quantum spin system without driving, Commun. Phys. $\textbf{8}$, 114 (2025).

\bibitem{Kongkhambut}
P. Kongkhambut, J. Skulte, L. Mathey, J. G. Cosme, A. Hemmerich, and H. Keßler, Observation of a continuous time crystal, Science $\textbf{377}$, 670 (2022).

\bibitem{Wu}
X. Wu, Z. Wang, F. Yang, R. Gao, C. Liang, M. K. Tey, X. Li, T. Pohl, and L. You, Dissipative time crystal in a strongly interacting Rydberg gas, Nat. Phys. $\textbf{20}$, 1389--1394 (2024).

\bibitem{Jiao}
Y. Jiao, W. Jiang, Y. Zhang, J. Bai, Y. He, H. Shen, J. Zhao, and S. Jia, Observation of multiple time crystals in a driven-dissipative system with Rydberg gas, Nat. Commun. $\textbf{16}$, 8767 (2025). DOI: 10.1038/s41467-025-64488-7.

\bibitem{Chen}
YH. Chen and X. Zhang, Realization of an inherent time crystal in a dissipative many-body system. Nat Commun $\textbf{14}$, 6161 (2023).

\bibitem{Greilich}
A. Greilich, N.E. Kopteva, A.N. Kamenskii, P.S. Sokolov, V.L. Korenev and M. Bayer, Robust continuous time crystal in an electron–nuclear spin system. Nat. Phys. $\textbf{20}$, 631–636 (2024).

\bibitem{Kuramoto}
Y. Kuramoto, Chemical Oscillations, Waves, and Turbulence (Springer Berlin Heidelberg, 1984)

\bibitem{Nakanishi2}
Y. Nakanishi, T. Sasamoto, Dissipative time crystals originating from parity-time symmetry, Phys. Rev. A $\textbf{107}$, L010201(2023).

\bibitem{Buca2}
B. Bu\v{c}a and T. Prosen, A note on symmetry reductions of the Lindblad equation: transport in constrained open spin chains, New J. Phys. $\textbf{14}$ 073007 (2012).

\bibitem{Kessler}
E. M. Kessler, G. Giedke, A. Imamoglu, S. F. Yelin, M. D. Lukin, and J. I. Cirac, Dissipative phase transition in a central spin system, Phys. Rev. A $\textbf{86}$, 012116 (2012).

\bibitem{Minganti}
F. Minganti, A. Biella, N. Bartolo, $\&$ C. Ciuti, Spectral theory of Liouvillians for dissipative phase transitions, Phys. Rev. A $\textbf{98}$, 042118 (2018).

\bibitem{Huber2}
J. Huber, P. Kirton, S. Rotter, $\&$ P. Rabl, Emergence of $\mathcal{PT}$-symmetry breaking in open quantum systems, SciPost Phys. $\textbf{9}$, 052 (2020).

\bibitem{Konotop}
V. V. Konotop, J. Yang, and D. A. Zezyulin, Nonlinear waves in $\mathcal{PT}$-symmetric systems, Rev. Mod. Phys. $\textbf{88}$, 035002 (2016).

\bibitem{Fruchart}
M. Fruchart, R. Hanai, P. B. Littlewood, and V. Vitelli, Nonreciprocal phase transitions, Nature (London) $\textbf{592}$, 363 (2021).

\bibitem{You}
Z. You, A. Baskaran, and M. C. Marchetti, Nonreciprocity as a Generic Route to Traveling States, Proc. Natl. Acad. Sci. U.S.A. $\textbf{117}$, 19767 (2020).

\bibitem{Saha}
S. Saha, J. Agudo-Canalejo, and R. Golestanian, Scalar Active Mixtures: The Nonreciprocal Cahn-Hilliard Model, Phys. Rev. X $\textbf{10}$, 041009 (2020).

\bibitem{Hanai}
R. Hanai, A. Edelman, Y. Ohashi, and P. B. Littlewood, Non-Hermitian phase transition from a polariton Bose-Einstein condensate to a photon laser, Phys. Rev. Lett. $\textbf{122}$, 185301(2019).

\bibitem{Hanai2}
R. Hanai and P. B. Littlewood, Critical fluctuations at a many-body exceptional point, Phys. Rev. Res. $\textbf{2}$, 033018(2020).

\bibitem{Suchanek}
T. Suchanek, K. Kroy, and S. A. M. Loos, Entropy production in the nonreciprocal Cahn-Hilliard model, Phys. Rev. E $\textbf{108}$, 064610 (2023).

\bibitem{Zelle}
C. P. Zelle, R. Daviet, A. Rosch, and S. Diehl, Universal phenomenology at critical exceptional points of nonequilibrium $O(N)$ models, Phys. Rev. X $\textbf{14}$, 021052 (2024).

\bibitem{Chiacchio}
E. I. Rodr\'{\i}guez Chiacchio, A. Nunnenkamp, and M. Brunelli, Nonreciprocal Dicke Model, Phys. Rev. Lett. $\textbf{131}$, 113602 (2023).

\bibitem{Nadolny}
T. Nadolny, C. Bruder, M. Brunelli, Nonreciprocal synchronization of active quantum spins, Phys. Rev. X $\textbf{15}$, 011010 (2025).

\bibitem{ptantipt}
As we have shown in the main text, DCTCs in open quantum systems can be understood as a $PT$ symmetric phase of the dynamical systems described by the nonlinear Schr\"odinger-type equation, $i\partial_t {\bf{m}}={\bf{f}}({\bf{m}})$, where the $PT$ symmetry is defined as $P{\bf{f}}^*({\bf{q}}(t))={\bf{f}}(P{\bf{q}}^*(t))$. Nonreciprocal phase transitions~\cite{Fruchart, You, Saha}, on the other hand, correspond to a spontaneous symmetry breaking of \textit{anti}-$PT$ symmetry, $P{\bf{f}}^*({\bf{q}}(t))=-{\bf{f}}(P{\bf{q}}^*(t))$. [Note that, in Ref.~\cite{Fruchart}, they called their transition $PT$ symmetry breaking due to the use of different conventions from our definition of $PT$ symmetry.] This difference arises due to the property that the order parameter dynamics of our open quantum systems are described by the form similar to the nonlinear Schr\"odinger equation, $i\partial_t {\bf{q}}={\bf{f}}({\bf{q}})$, while those considered in active matter systems
are overdamped, $\partial_t {\bf{q}}={\bf{g}}({\bf{q}})$, where the factor ``$i$'' in front of the time derivative on the left-hand side is missing compared to the open quantum system analog.

\bibitem{Lindblad}
G. Lindblad, On the generators of quantum dynamical semigroups, Commun. Math. Phys. $\textbf{48}$, 119 (1976).

\bibitem{GKS}
V. Gorini, A. Kossakowski, E. C. G. Sudarshan, Completely positive dynamical semi-groups of $N$-level systems, J. Math. Phys. $\textbf{17}$, 821 (1976).

\bibitem{Huber1}
J. Huber, P. Kirton, $\&$ P. Rabl, Nonequilibrium magnetic phases in spin lattices with gain and loss, Phys. Rev. A $\textbf{102}$, 012219 (2020).

\bibitem{Nakanishi1}
Y. Nakanishi, T. Sasamoto, $\mathcal{PT}$ phase transition in open quantum systems with Lindblad dynamics, Phys. Rev. A $\textbf{105}$, 022219 (2022).

\bibitem{Prosen3}
T. Prosen, $\mathbb{PT}$-symmetric quantum Liouvillean dynamics, Phys. Rev. Lett. $\textbf{109}$, 090404 (2012).

\bibitem{Prosen4}
T. Prosen, Generic examples of $\mathbb{PT}$-symmetric qubit (spin-1/2) Liouvillian dynamics, Phys. Rev. A $\textbf{86}$, 044103 (2012).

\bibitem{Huybrechts}
D. Huybrechts, F. Minganti, F. Nori, M. Wouters, $\&$ N. Shammah, Validity of mean-field theory in a dissipative critical system: Liouvillian gap, $\mathbb{PT}$-symmetric antigap, and permutational symmetry in the XYZ model, Phys. Rev. B $\textbf{101}$, 214302 (2020).

\bibitem{Sa}
L. S\'{a}, P. Ribeiro, and T. Prosen, Symmetry classification of many-body Lindbladians: Tenfold way and beyond, Phys. Rev. X $\textbf{13}$, 031019 (2023).

\bibitem{App}
The $\mathcal{PT}$ symmetry for a shifted Lindbladian $\hat{\mathcal{L}}^\prime:=\hat{\mathcal{L}}-\alpha\hat{1}$ with $\alpha:=\text{Tr}\hat{\mathcal{L}}/\text{Tr}\hat{1}$ is defined as $\hat{\mathcal{P}}(\hat{\mathcal{L}}^\prime)^\dagger(\hat{\mathcal{P}})^{-1}=-\hat{\mathcal{L}}^\prime$ \cite{Prosen3}. Here, $\hat{\mathcal{P}}$ is a parity superoperator.

\bibitem{supsup}
See the Supplemental Materials below for details of the Lindbladian $\mathcal{PT}$ symmetry, mean-field derivations, stability analysis, and model calculations, which includes Refs.~\cite{opticsbooks,Note4}.

\bibitem{opticsbooks}
C. C. Gerry and P. L. Knight, \textit{Introductory Quantum Optics} (Cambridge University Press, Cambridge, England, 2005).

\bibitem{Note4}
Precisely speaking, L-$\mathcal{PT}$ symmetry [Eq.~(1) in the main text] is satisfied when the Lindbladian is invariant under the transformation, $\{l_\mu\}\to\{l_\mu+c_\mu I\}$ and $h\to h+i\sum_\mu(c_\mu^*l_\mu-c_\mu l_\mu^\dagger)$ ($c_\mu\in\mathbb{C}$), which is slightly more generic than the transformation~(B.18). One can readily check that Eq.~(B.19) is still satisfied under such a constraint.

\bibitem{Souza}
L. S. Souza, L. F. dos Prazeres, and F. Iemini, Sufficient condition for gapless spin-boson Lindbladians, and its connection to dissipative time crystals, Phys. Rev. Lett. $\textbf{130}$, 180401 (2023).

\bibitem{Carollo}
F. Carollo and I. Lesanovsky, Exactness of mean-field equations for open Dicke models with an application to pattern retrieval dynamics, Phys. Rev. Lett. $\textbf{126}$, 230601 (2021).

\bibitem{Strogatz}
S. Strogatz, Nonlinear Dynamics and Chaos: With Applications to Physics, Biology, Chemistry, and Engineering (CRC Press, Boca Raton, FL, 2018).

\bibitem{Note1}
We remark that persistent oscillations induced by L-$\mathcal{PT}$ symmetry are fundamentally distinct from those induced by decoherence-free subspaces \cite{Lidar} or strong dynamical symmetries \cite{Buca, Booker}. In our case, the period of persistent oscillations depends on dissipation strength, while in the latter, it is determined solely by the Hamiltonian.

\bibitem{Lidar}
D. A. Lidar, I. L. Chuang, and K. B. Whaley, Decoherence-Free Subspaces for Quantum Computation, Phys. Rev. Lett. $\textbf{81}$, 2594 (1998).

\bibitem{Buca}
B. Bu\v{c}a, J. Tindall, and D. Jaksch, Non-stationary coherent quantum many-body dynamics through dissipation, Nat. Commun. $\textbf{10}$, 1730 (2019).

\bibitem{Bender}
C. M. Bender and S. Boettcher, Real spectra in non-Hermitian Hamiltonians having $\mathcal{PT}$ symmetry, Phys. Rev. Lett. $\textbf{80}$, 5243 (1998); C. M. Bender, S. Boettcher, and P. N. Meisinger, PT-symmetric quantum mechanics, J. Math. Phys. $\textbf{40}$, 2201 (1999).

\bibitem{Bender2}
C. M. Bender, Introduction to PT-symmetric quantum theory, Contemp. Phys. $\textbf{46}$, 277 (2005); Making sense of non Hermitian Hamiltonians, Rep. Prog. Phys. $\textbf{70}$, 947 (2007).

\bibitem{MostafazadehA1}
A. Mostafazadeh, Pseudo-Hermiticity versus PT symmetry: The necessary condition for the reality of the spectrum of a non-Hermitian Hamiltonian, J. Math. Phys. $\textbf{43}$, 205 (2002).

\bibitem{Note2}
The DCTCs originating from L-$\mathcal{PT}$ symmetry cannot emerge in the effective non-Hermitian Hamiltonian obtained by neglecting quantum jump terms in the GKSL equation, $H_{\rm eff}:=H-i\sum_\mu L_\mu^\dagger L_\mu$, since it cannot have (active) PT symmetry $[H_{\rm eff},PT]\neq 0$.

\bibitem{Ribeiro}
P. Ribeiro and T. Prosen, Integrable Quantum Dynamics of Open Collective Spin Models, Phys. Rev. Lett. $\textbf{122}$, 010401 (2019).

\bibitem{Carmichael}
H. J. Carmichael, Analytical and numerical results for the steady state in cooperative resonance fluorescence, J. Phys. B $\textbf{13}$, 3551 (1980).

\bibitem{Hannukainen}
J. Hannukainen and J. Larson, Dissipation-driven quantum phase transitions and symmetry breaking, Phys. Rev. A $\textbf{98}$, 042113 (2018).

\bibitem{Ferreira}
J. S. Ferreira and P. Ribeiro, Lipkin-Meshkov-Glick model with Markovian dissipation: A description of a collective spin on a metallic surface, Phys. Rev. B $\textbf{100}$, 184422 (2019).

\bibitem{Pires}
A. S. T. Pires, Theoretical tools for spin models in magnetic systems (IOP, Bristol, UK, 2021).

\bibitem{Baumann}
K. Baumann, C. Guerlin, F. Brennecke, and T. Esslinger, Dicke quantum phase transition with a superfluid gas in an optical cavity, Nature (London) $\textbf{464}$, 1301 (2010).

\bibitem{Baumann2}
K. Baumann, R. Mottl, F. Brennecke, and T. Esslinger, Exploring Symmetry Breaking at the Dicke Quantum Phase Transition, Phys. Rev. Lett. $\textbf{107}$, 140402 (2011).

\bibitem{Zhang}
J. Zhang, G. Pagano, P. W. Hess, A. Kyprianidis, P. Becker, H. Kaplan, A. V. Gorshkov, Z.-X. Gong, and C. Monroe, Observation of a many-body dynamical phase transition with a 53-qubit quantum simulator, Nature (London) $\textbf{551}$, 601 (2017).

\bibitem{Angerer}
A. Angerer, K. Streltsov, T. Astner, S. Putz, H. Sumiya, S. Onoda, J. Isoya, W. J. Munro, K. Nemoto, J. Schmiedmayer, and J. Majer, Superradiant emission from colour centres in diamond, Nat. Phys. $\textbf{14}$, 1168 (2018).

\bibitem{Ferioli}
G. Ferioli, A. Glicenstein, I. Ferrier-Barbut, and A. Browaeys, A non-equilibrium superradiant phase transition in free space, Nat. Phys. $\textbf{19}$, 1345 (2023).

\bibitem{Note3}
For finite $S$, the relaxation time $\tau _S$ of the oscillating modes of DCTCs in one-collective spin systems is known to scale with $S$ \cite {Iemini} or $\log S$ \cite {Piccitto}. This scaling suggests that in large spin systems, DCTCs can be observed if their frequency $\omega $ is sufficiently high $\omega \gg 1/ \tau _S$.

\bibitem{Sup2}
The DDM (i.e. $\omega=0$) is a subtle case where a CEP appears, but its effects cannot be observed. This is because the fluctuation vector orthogonal to the zero excitation mode coincides with the normal vector, but this cannot be excited because $S$ has to be conserved (see SM Sec. D.2 \cite{supsup}).

\bibitem{Lipkin}
H. Lipkin, N. Meshkov, and A. Glick, Validity of many-body approximation methods for a solvable model: (I). Exact solutions and perturbation theory, Nucl. Phys. $\textbf{62}$, 188 (1965).

\bibitem{Lee}
T. E. Lee, C.-K. Chan, and S. F. Yelin, Dissipative phase transitions: Independent versus collective decay and spin squeezing, Phys. Rev. A $\textbf{90}$, 052109 (2014).

\bibitem{Qutip}
J. R. Johansson, P. D. Nation, and F. Nori, QuTiP: An open-source Python framework for the dynamics of open quantum systems, Comput. Phys. Commun. $\textbf{183}$, 1760 (2012).

\bibitem{NakanishiData2026}
https://zenodo.org/records/20151882.

\edef\@currentlabel{76}
\label{LastBibItem}
\makeatother
\end{thebibliography}
\end{document}